\documentclass[structabstract]{aa}  
\usepackage{graphicx}
\usepackage{amsmath,amssymb}
	\DeclareMathOperator{\magn}{mag}
\usepackage{txfonts}
\usepackage{natbib}
\usepackage{multirow}
\begin{document}
   \title{Improved achromatization of phase mask coronagraphs using colored apodization}

   \author{M. N'Diaye\inst{\ref{LAM},\ref{IAUNAM},\ref{GIS-PHASE}}
     \and
     K. Dohlen\inst{\ref{LAM},\ref{GIS-PHASE}}
     \and
     S. Cuevas\inst{\ref{IAUNAM}}
     \and
     R. Soummer\inst{\ref{STScI}}
     \and
     C. S\'anchez-P\'erez\inst{\ref{CCADET}}
     \and
     F. Zamkotsian\inst{\ref{LAM}}
   }

   \institute{Laboratoire d'Astrophysique de Marseille, UMR6110, OAMP, CNRS/Universit\'e de Provence\\38 rue Fr\'ed\'eric Joliot-Curie, 13388 Marseille cedex 13, France\\
             \email{mamadou.ndiaye@oamp.fr, kjetil.dohlen@oamp.fr}\label{LAM}
         \and
             Instituto de Astronom\'\i{}a, Universidad Nacional Aut\'onoma de M\'exico (UNAM)\\Apartado Postal 70-264 Ciudad Universitaria, 04510 M\'exico D.F., Mexico\label{IAUNAM}
         \and
						 Space Telescope Science Institute, 3700 San Martin Drive, Baltimore, MD 21218, USA\label{STScI}
				 \and
				 		 Centro de Ciencias Aplicadas y Desarrollo Tecnol\'ogico, UNAM\\Apartado Postal 70-186 Ciudad Universitaria, 04510 M\'exico D.F., Mexico\label{CCADET}
				 \and 
				     Groupement d'Int\'er\^et Scientifique PHASE (Partenariat Haute R\'esolution Angulaire Sol-Espace\label{GIS-PHASE})
           }

   \date{Received ********; accepted ********}

 
  \abstract
   {For direct imaging of exoplanets, a stellar coronagraph helps to remove the image of an observed bright star by attenuating the diffraction effects caused by the telescope aperture of diameter $D$. The Dual Zone Phase Mask (DZPM) coronagraph constitutes a promising concept since it theoretically offers a small inner working angle ($IWA \sim \lambda_0/D$ where $\lambda_0$ denotes the central wavelength of the spectral range $\Delta\lambda$), good achromaticity and high starlight rejection, typically reaching a 10$^6$ contrast at 5\,$\lambda_0/D$ from the star over a spectral bandwidth $\Delta\lambda/\lambda_0$ of 25\% (similar to H-band). This last value proves to be encouraging for broadband imaging of young and warm Jupiter-like planets.}
   {Contrast levels higher than $10^6$ are however required for the observation of older and/or less massive companions over a finite spectral bandwidth. An achromatization improvement of the DZPM coronagraph is therefore mandatory to reach such performance.}
   {In its design, the DZPM coronagraph uses a grey (or achromatic) apodization. We propose to replace it by a colored apodization to increase the performance of this coronagraphic system over a large spectral range. This innovative concept, called Colored Apodizer Phase Mask (CAPM) coronagraph, is defined with some design parameters optimized to reach the best contrast in the exoplanet search area. Once this done, we study the performance of the CAPM coronagraph in the presence of different errors to evaluate the sensitivity of our concept.}
   {A 2.5\,$\,\magn$ contrast gain is estimated from the performance provided by the CAPM coronagraph with respect to that of the DZPM coronagraph. A $2.2 \cdot 10^{-8}$ intensity level at 5\,$\lambda_0/D$ separation is then theoretically achieved with the CAPM coronagraph in the presence of a clear circular aperture and a 25\% bandwidth. In addition, our studies show that our concept is less sensitive to low than high-order aberrations for a given value of rms wavefront errors.}
   {}

   \keywords{Instrumentation: high angular resolution -- Techniques: high angular resolution -- Telescopes -- Methods: numerical}

   \titlerunning{Phase mask coronagraphs using colored apodization}
   \maketitle

%

\section{Introduction}\label{sec:intro}  
\subsection{Astronomical context}
Recently, the direct imaging of several exoplanets have been made possible \citep{2008Sci...322.1345K,2008Sci...322.1348M,2010Sci...329...57L} owing to their youth (tens of Myr), their high mass (a few Jupiter masses) or their wide apparent distance from their host star (larger than 1 arcsecond). For imaging older, less massive or less separated planetary companions while enabling their spectroscopic observations, achromatic high contrast imaging techniques are required. However, large broadband imaging of these exoplanets is extraordinarily challenging in the visible and near-infrared bands
because of the large flux ratio ($10^7$ to $10^{10}$) and the small angular separation (less than 1 arcsecond) between them and their host star \citep{2010exop.book..111T}.\\
Such detection requires the combination of different methods, such as extreme adaptive optics (XAO), stellar coronagraphy and post processing techniques, to enhance the exoplanet image \citep{2009ARA&A..47..253O}. XAO systems on ground-based telescopes \citep{2008SPIE.7015E..31M,2010SPIE.7736E..13S,2010SPIE.7736E..71G,2011PASP..123...74H} will compensate the effects of the atmospheric turbulence to provide high angular resolution images at very high Strehl ratio. A stellar coronagraph \citep{2006ApJS..167...81G} will help to mostly suppress the diffraction effects due to the telescope aperture. In addition, post-processing techniques \citep[e.g.][]{2000PASP..112...91M,2002ApJ...578..543S,2006ApJ...641..556M,2007ApJ...660..770L,2008OExpr..1618406M,2008A&A...489.1345V} will allow us to remove the speckles caused by the XAO residual aberrations and the slow variations of the physical conditions of a telescope.\\ The current limits of all these methods need to be pushed further to expect imaging the exoplanets mentioned above. In particular, increasing the performance of the current coronagraphs is mandatory for reaching an excellent polychromatic removal of the star diffraction pattern and image the faintest substellar mass companions in white light. The resulting, innovative coronagraphs will find their place in the exoplanet imagers planned to follow the upcoming Palomar P1640 \citep{2011PASP..123...74H}, VLT-SPHERE \citep{2008SPIE.7014E..41B}, Gemini Planet Imager \citep{2008SPIE.7015E..31M} and Subaru HiCIAO \citep{2008SPIE.7014E..42H}: e.g. EPICS \citep{2008SPIE.7015E..46K} for the European Extremely Large Telescope (E-ELT) on the ground or possible 1-2\,m off-axis telescopes in space \citep{2010SPIE.7731E..65G,2010SPIE.7731E..67T,2010SPIE.7731E..68G}. 

\subsection{Broadband capabilities of coronagraphs}
Many coronagraphic designs have been proposed these past few years; most of them are reported in \citet{2006ApJS..167...81G}. We are here particularly interested in broadband behavior of coronagraphs and although a complete review of the broadband performance of all the existing coronagraphs is not within the scope of this paper, we provide here some numbers from the literature allowing to situate the presented work in the context of the state of the art. $\lambda_0$ and $D$ denote the central wavelength of the spectral range $\Delta\lambda$ and the telescope aperture diameter respectively.\\
The Achromatic Interfero Coronagraph \citep{1996CRASB.322..265G} was the first attempt to make a concept able to image faint substellar mass companions close to the star in white light. This device can theoretically offer an intensity level of $1.6\cdot 10^{-5}$ at a distance of $\lambda_0/D$ from the star over a bandwidth ($\Delta\lambda/\lambda_0$) of 18\% \citep{2000A&AS..141..319B}. The Apodized Pupil Lyot coronagraph \citep{2002A&A...389..334A,2003A&A...397.1161S}, a more recent concept, is one of the most popular designs, adopted for the forthcoming generation of exoplanet imagers (e.g. VLT-SPHERE, Gemini Planet Imager, Palomar P1640). It can theoretically reach an intensity level lower than $10^{-7}$ at 5\,$\lambda_0/D$ over a bandwidth of 20\%, in the presence of a telescope aperture with central obstruction \citep{2005PASP..117.1012A, 2005ApJ...618L.161S,2011ApJ...729..144S}. Band-limited coronagraphs \citep{2002ApJ...570..900K,2005ApJ...628..466K} also represent encouraging solutions to produce very high contrast performance over large spectral bands. Recent studies predict for instance that these concepts using hybrid metal-dielectric masks and combined with a deformable mirror, set to generate a dark half field in the image plane, can provide an intensity level of $4.9\cdot 10^{-10}$ at 3.5\,$\lambda_0/D$ within this dark region, over a bandwidth of 20\% \citep{2008SPIE.7010E.107M}. The Phase Induced Amplitude Apodization \citep[PIAA,][]{2003A&A...404..379G} and its avatars \citep{2005ApJ...622..744G,2006ApJ...639.1129M,2006ApJ...644.1246P,2010ApJS..190..220G} constitute some of the most promising solutions for Earth-like planet detection. Indeed, an intensity level of $10^{-10}$ at 1.5\,$\lambda_0/D$ over a 21\% bandwidth is predicted with an hybrid coronagraph composed of PIAA with a classical apodization \citep{2006ApJ...644.1246P}.\\
Stellar coronagraphs using a phase mask in the focal plane represent some encouraging solutions to achieve the requested contrast levels, either in the form of sectorized masks (Four Quadrant Phase Mask [\citealp{2000PASP..112.1479R}] and Eight Octant Phase Mask [\citealp{2008PASP..120.1112M}]) or circular masks \citep{1997PASP..109..815R,2003A&A...403..369S}. These systems are characterized by their very small inner working angles (IWAs) and good starlight rejection capabilities.\\
Achromatization of these phase mask coronagraphs has been intensively studied in the past few years to improve the capability of these concepts to work in broad spectral bands. For sectorized phase masks, the use of achromatic half-wave plates leads to intensity levels of $5 \cdot 10^{-5}$ at 2.5\,$\lambda_0/D$ from the star over a 20\% bandwidth \citep{2006A&A...448..801M}.\\ 
Derived from the FQPM concept, the vector vortex coronagraph \citep{2005ApJ...633.1191M,2010SPIE.7739E..33M} also provides achromatic solutions. Using Liquid Crystal Polymers (LCP) for the manufacturing of such phase mask, intensity levels lower than $1.4 \cdot 10^{-7}$ at 3\,$\lambda_0/D$ for a 15\% bandwidth can theoretically be reached \citep{2005ApJ...633.1191M}, and further improvements are expected using a 3-layer LCP structure \citep{2010SPIE.7739E..33M}.\\
Achromatization of circular phase mask coronagraphs by the use of dual zone phase structures has also been studied \citep{2003A&A...403..369S}, leading to intensity levels of $10^{-6}$ at 3\,$\lambda/D$ for 20\% bandwidth. We report on further improvements of this concept in the following.

\subsection{Circular phase mask coronagraphs}
Focus is made here on stellar coronagraphs with circular axi-symmetrical phase masks \citep{1997PASP..109..815R, 2003A&A...403..369S}. This is a promising concept, well suited for any arbitrary aperture (clear aperture, centrally obstructed circular pupil, diluted pupil...) and free from blind axes in the coronagraphic images. \citet{1997PASP..109..815R} proposed the first design, referred here as Roddier \& Roddier Phase Mask (RRPM), which was later improved by adding an entrance pupil apodizer \citep{2000SPIE.4006..377G,2003A&A...397.1161S}. The corresponding system, called apodized pupil Roddier \& Roddier phase mask (ARPM) coronagraph, can theoretically achieve perfect starlight extinction at a single wavelength. Unfortunately, its performance decreases strongly in broadband observation. This loss of coronagraphic performance comes from chromatism effects related to the inherent properties of the RRPM.\\ 
\citet{2003A&A...403..369S} proposed an improvement of the ARPM coronagraph to overcome the problem of chromaticity, replacing the simple phase mask of the Roddier coronagraph with a double phase mask, named Dual Zone Phase Mask (DZPM). With this innovative coronagraph which also includes pupil apodization, they showed that an intensity level of $\sim 10^{-6}$ can theoretically be reached at a 3\,$\lambda_0/D$ angular distance from the star. These promising results were achieved both for a clear circular aperture and for an entrance pupil with 14\% central obscuration ratio, in both cases considering a 20\% bandwidth. Although encouraging, the performance of this coronagraphic system needs to be enlarged so as to realize spectro-imaging observations of the faintest close-in companions with this concept. The DZPM coronagraph has been originally designed considering a grey (or achromatic) entrance pupil apodization. This component then applies the same transmission function to the incident beam over the whole spectral bandwidth of study. The apodization shape is possibly not optimal at all the wavelengths of study.

\subsection{Objectives of the paper}
The aim of this paper is twofold. First of all, we investigate the possibility of using a colored (or chromatic) instead of a grey apodization to improve the contrast gain delivered by the DZPM coronagraph over the whole spectral bandwidth. This innovative concept will be named Colored Apodizer Phase Mask (CAPM) coronagraph. Secondly, once the apodization parameters optimized, we propose to analyze the sensitivity of the CAPM coronagraph by estimating the impact of different effects which may alter its performance in a realistic case.\\
In Section \ref{sec:CAPM}, we briefly review the principle of the DZPM coronagraph and state the possibility of upgrading this concept with the use of a colored apodization. The methodology, applied to optimize numerically the parameters of the CAPM coronagraph, is thereafter described in Section \ref{sec:methodology}. We present the theoretical performance achieved by this coronagraphic device for different spectral bandwidths and a clear circular aperture in Section \ref{sec:performance}. In addition, manufacturing aspects are addressed in Section \ref{sec:design} with the development of a physical design for the colored apodization and the performance study for the CAPM coronagraph in the presence of this component. The sensitivity of the CAPM coronagraph to different errors is thereafter analyzed in Section \ref{sec:sensitivity}.

\section{The CAPM coronagraph}\label{sec:CAPM}
The principle of the DZPM coronagraph is recalled in the following before introducing our proposal, the CAPM coronagraph and its colored apodization. We then expose the formalism of this latter. 

\subsection{Principle of the DZPM coronagraph}
\begin{figure}[!ht]
\centering
\resizebox{\hsize}{!}{\includegraphics{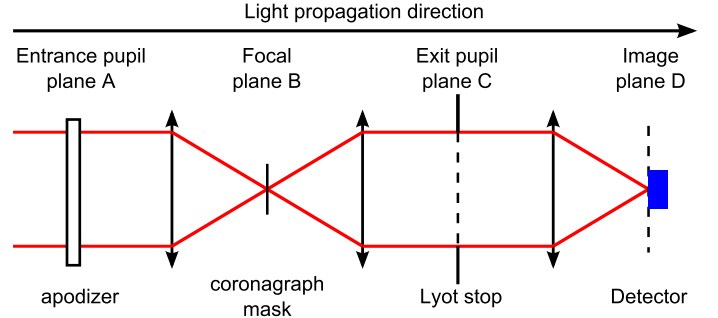}}
\caption{Scheme of the coronagraphic layout.} 
\label{fig:apod_coronagraph_layout}
\end{figure}

\begin{figure}[!ht]
\centering
\resizebox{\hsize}{!}{\includegraphics{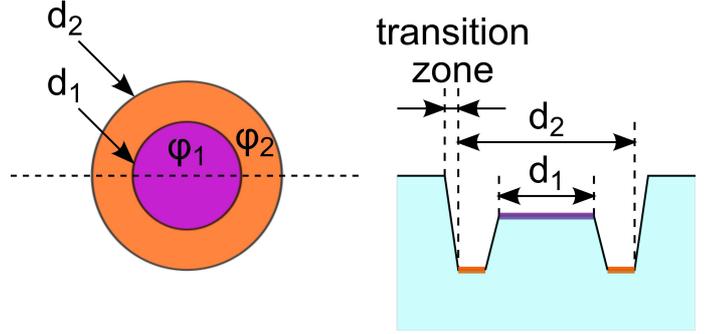}}
\caption{Schematic representation of the DZPM. The dimensions are not at scale in this drawing.} 
\label{fig:DZPM_drawing}
\end{figure}

An illustration of the DZPM coronagraphic layout is given in Figure \ref{fig:apod_coronagraph_layout}, detailing the typical four planes of Lyot style coronagraphs. The DZPM coronagraph combines an apodization in the entrance pupil plane A, a DZPM in the following focal plane B and a Lyot stop in the re-imaged pupil plane C. The coronagraphic image forms in the final image plane D in which a detector is located. The optical design is such that the complex amplitudes of the electric field in two successive planes are related one to each other by a Fourier transform operation.\\
The DZPM is designed as a circular phase disk of diameter $d_1$ surrounded by an annular phase ring of diameter $d_2$, see Figure \ref{fig:DZPM_drawing}. Each diameter is equal to a very precise fraction of the size of the Airy disk. The inner and outer parts of the mask introduce some phase shifts $\varphi_1$ and $\varphi_2$ respectively, to the incoming wavefront in the focal plane B. As the wavelength increases, $\varphi_1$ and $\varphi_2$ decrease and the stellar diffraction pattern expands. This combination of effects leads to a decrease of the relative flux passing through the inner zone, while the flux passing through the annular zone increases. At a given wavelength, the complex amplitude of the field in the relayed pupil plane C corresponds to the result of the interference between the direct wave with the waves delayed by the two phase shift zones of the DZPM. This sum of wavefronts varies with the wavelength since $\varphi_1$, $\varphi_2$ and the areas of the diffraction pattern covered by the mask also evolve with the wavelength. With an adequate choice of the DZPM parameters, the recombination of these three wavefronts in plane C generates destructive interferences inside the geometric pupil for a wide range of wavelengths. The light is rejected outside of this relayed pupil and blocked by a Lyot stop, leading to a broadband suppression of the star image at the detector in plane D.\\
In addition, the presence of an entrance pupil apodization in plane A allows us to match well the direct wave with the waves diffracted by the DZPM. It results in improved destructive interference over the relayed pupil. A lower amount of the residual starlight is then found inside this pupil and therefore, a better starlight extinction is observed in the final image plane D. As was shown by \citet{2003A&A...403..369S}, further improvements can be achieved by adding a slight defocus of the coronagraph mask, which is equivalent to allocate a complex term in our apodization.

\subsection{Replacement of the grey by a colored apodization}
A complex apodization in the coronagraphic scheme allows the direct wave and the waves diffracted by the DZPM to match well within the geometric pupil image and interfere destructively over it. In their design, \citet{2003A&A...403..369S} considered a grey version of the complex apodization. Since the waves diffracted by the mask evolve with the wavelength, they do not match well the direct wave at all the wavelengths. Ideally, a complex apodization has to be found for each wavelength to obtain an achromatic nulling recombination of the direct with the diffracted waves.\\
The design improvement proposed here from the DZPM coronagraph is similar to the approach developed by \citet{2005PASP..117.1012A} for the Prolate apodized Lyot coronagraph. The idea consists in replacing the grey by a colored apodization in our coronagraphic system. This colored apodization constitutes an amplitude apodization and hence, intensity apodization variable with the wavelength. This change in apodization leads to the design of CAPM coronagraph which combines a colored apodization, a DZPM in the following focal plane and a Lyot stop in the relayed pupil plane. With this innovative coronagraphic system, we expect to reach contrast levels larger than those achieved by the DZPM coronagraph over a finite spectral bandwidth.

\subsection{Formalism}
In the following, the formalism common to DZPM and CAPM coronagraphs is given. For the sake of clarity, we omit the position vector $\textbf{r}$, its modulus $r$ and the wavelength $\lambda$ in the equations below. $\mathcal{F}$ symbolizes the Fourier transform operator in which we include the Fourier Optics scaling factor $1/\lambda f$, with $f$ the telescope focal length \citep{1996ifo..book.....G}.\\
The complex amplitude of the field $\Psi_A$ at the aperture is given by:
\begin{equation}
\Psi_A=P\,\Phi\,,
\label{eq:ampltude_A}
\end{equation}
in which $P$ defines the telescope aperture shape whereas $\Phi$ denotes the complex apodization.\\
The DZPM is located in the following focal plane B. Its amplitude transmission function $t$ can be written as:
\begin{equation}
t=1 - (e^{i\varphi_2}- e^{i\varphi_1})\,M_1 - (1- e^{i\varphi_2})\,M_2\,,
\label{eq:transmission_DZPM}
\end{equation}
in which M$_1$ and M$_2$ define the top-hat functions of the inner and outer parts of the DZPM; they are equal to 1 for $|r|<d_1/2$ and $|r|<d_2/2$ respectively and 0 otherwise. Let us remind that $\varphi_1$ and $\varphi_2$ represent the phase shifts introduced by the mask in its inner and outer parts respectively, at a given wavelength.\\
The complex amplitude of the field $\Psi_B$ after the mask is then given by:
\begin{equation}
\Psi_B=\left [1 - (e^{i\varphi_2}- e^{i\varphi_1})\,M_1\, - (1- e^{i\varphi_2})\,M_2 \right ]\,\mathcal{F}[\Psi_A]\,.
\label{eq:amplitude_B}
\end{equation}
The DZPM and CAPM coronagraphs belong to the class of the Lyot-style coronagraphs which consist in a succession of binary filters (pupil, finite-size focal plane mask, Lyot stop). We then adopt the semi-analytical approach suggested by \citet{2007OExpr..1515935S} in the case of a Lyot-style coronagraph for a better understanding of the interference effects occurring in the pupil plane C. The complex amplitude after the Lyot stop $\Psi_C$ is expressed as follows:
\begin{equation}
\begin{split}
\Psi_C = & \,\Psi_A\,L \\
& - (e^{i\varphi_2}- e^{i\varphi_1})\,\mathcal{F}\left [\mathcal{F}[\Psi_A]M_1 \right ]\,L \\
& - (1- e^{i\varphi_2})\,\mathcal{F} \left [\mathcal{F}[\Psi_A]M_2 \right ]\,L \,,
\label{eq:ampltude_C}
\end{split}
\end{equation}
in which L denotes the index function of the Lyot stop. $\Psi_C$ is the sum of three terms which represent the direct wave $\Psi_A$ and the sum of the waves diffracted by the inner and outer parts of the DZPM. As mentioned above, perfect starlight cancellation is achieved if the sum of the waves diffracted by the DZPM matches perfectly the direct wave inside the geometric pupil.\\
The coronagraphic amplitude $\Psi_D$ in the image plane D is finally obtained with:
\begin{equation}
\Psi_D=\mathcal{F}[\Psi_C] \,.
\label{eq:ampltude_D}
\end{equation}
The corresponding intensity $I_{D}$ of the polychromatic coronagraphic image is given by: 
\begin{equation}
I_{D}=\frac{1}{\Delta\lambda}\int_{\lambda_0-\Delta\lambda/2}^{\lambda_0+\Delta\lambda/2} |\Psi_D|^2\,d\lambda \,.
\label{eq:intensity_D}
\end{equation}
We recall that $\Delta\lambda$ defines the spectral range of study centered at the wavelength $\lambda_0$.

\section{Methodology}\label{sec:methodology}
In this section, we describe the methodology used to realize our numerical simulations of the CAPM coronagraph. We also expose the optimization criteria to determine the parameters of this latter.

\subsection{Parameters of the CAPM coronagraph}
Several parameters are involved in the CAPM coronagraph. $d_1$, $d_2$ and the phase steps, expressed in optical path differences OPD$_1$ and OPD$_2$, correspond to the characteristics related to the DZPM. The other parameters refer to the complex apodization $\Phi$ which can be decomposed into the product of an amplitude and a phase apodizations $\Phi_a$ and $\Phi_w$ respectively:
\begin{equation}
\Phi=\Phi_a\cdot\Phi_w \,.
\label{eq:decomp_apodization}
\end{equation}
Following the indications of \citet{2003A&A...403..369S}, the transmission function chosen for $\Phi_a$ is a radial symmetric fourth order polynomial while $\Phi_w$ is in the form of a defocus of the mask. This amplitude transmission function finds its maximum value at the pupil center in the case of a clear aperture and at the edge of the central obscuration for a centrally obstructed circular aperture. For a given central obscuration ratio $\eta$ (null for a clear aperture), the amplitude apodization $\Phi_a$ can be written as:
\begin{equation}
\Phi_a(\textbf{r}, \lambda)= 1 + \omega_{1, \lambda}\left(r^2-\frac{\eta^2}{4}\right ) + \omega_{2, \lambda}\left (r^4-\frac{\eta^4}{16}\right ) \,,\\
\label{eq:ampl_apodization}
\end{equation} 
where $\omega_{1, \lambda}$ and $\omega_{2, \lambda}$ denote the colored parameters related to the amplitude apodization $\Phi_a$. In addition, the phase apodization $\Phi_w$ is expressed as follows:
\begin{equation}
\Phi_w(\textbf{r}, \lambda)= \exp(2i\pi\beta_{\lambda} r^2 \lambda_0/\lambda)\,,
\label{eq:phase_apodization}
\end{equation} 
where $\beta_{\lambda}$ is a chromatic parameter associated with the phase apodization. The chromatic longitudinal defocus $\Delta z$ is related to $\beta_{\lambda}$ by the following expression:
\begin{equation}
\Delta z(\lambda)= 2 \beta_{\lambda} f^2 \lambda\,,
\label{eq:Delta_z}
\end{equation} 
where $f$ denotes the optical focal length of the system. Manufacturing aspects on the chromatic defocus are discussed below in Section \ref{sec:design}. As mentioned above, $\omega_{1, \lambda}$, $\omega_{2, \lambda}$ and $\beta_{\lambda}$ are chromatic parameters since we deal here with a colored apodization. They will be optimized at each wavelength of the spectral bandwidth with the criteria given below. Let us note that the previous expressions of apodization are still valid for the DZPM coronagraph, with $\omega_{1, \lambda}$, $\omega_{2, \lambda}$ and $\beta_{\lambda}$ remaining constant over the spectral bandwidth.

\subsection{Numerical simulations using ultrafast Fourier transforms}
We detail here the computation of the amplitudes $\Psi_C$ and $\Psi_D$, respectively based on a semi-analytical approach and a novel method providing fast computation of high resolution Fourier transform.

\subsubsection{Computation of the amplitude in plane C}
The complex amplitude $\Psi_C$ in pupil plane C is computed following Eq.(\ref{eq:ampltude_C}). This semi-analytical expression is advantageous in terms of computation time since Fourier transform operations are only required within the mask area of width $\sim \lambda_0/D$ in the focal plane B and within the Lyot stop area in plane C. No zero-padding is required in this case so we can use arrays of width equal to the mask diameter. We apply the algorithm, proposed by \citet{2007OExpr..1515935S}, based on the matrix Fourier transform (MFT) to compute these truncated Fourier transform and evaluate $\Psi_C$.

\subsubsection{Computation of the amplitude in plane D}
\begin{figure}[!ht]
\centering
\resizebox{\hsize}{!}{\includegraphics{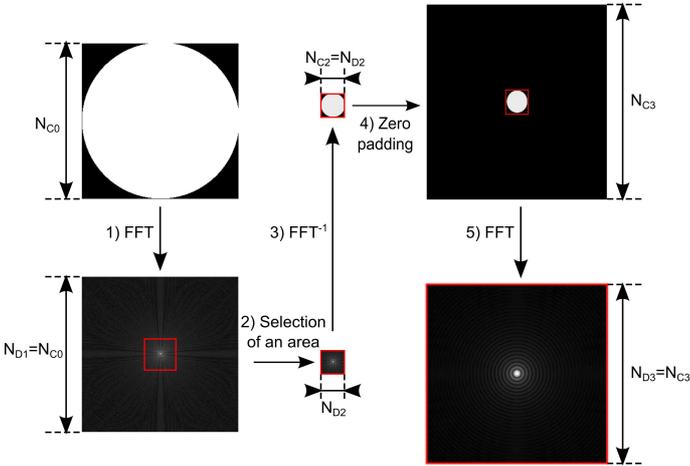}}
\caption{Schematic illustration of the method used for the computation of $\Psi_D$.} 
\label{fig:FFTs_method_scheme}
\end{figure}

The complex amplitude $\Psi_D$ in image plane D is obtained with a Fourier transform of the complex amplitude $\Psi_C$, see Eq.(\ref{eq:ampltude_D}). We want to evaluate this complex amplitude over a large field of view (FoV), typically from 50 to 200\,$\lambda_0/D$ width with good sampling. This can be seen as a truncated Fourier transform within the area of interest (here, FoV) like in the case of $\Psi_B$ and the MFT can be used again. However, the computation proves to be slow since an array much larger than that used in plane B is involved. While this increase in computation time is not dramatic in the monochromatic case, it becomes important when we consider a broadband image for which several monochromatic images are computed and summed.\\
For fast computation of $\Psi_D$ from $\Psi_C$, we apply a novel method based on a succession of Fast Fourier Transforms (FFTs) instead of the MFT. By this method, already experimented in the context of James Webb Space Telescope PSF calculation \citep{2004SPIE.5487..635Z}, the diameter in pixels of the geometric pupil in plane C is assumed to be equal to $N_{C0}$. $\Psi_C$ is represented in an array $C_0$ with dimensions $N_{C0} \times N_{C0}$. In the following, we describe the successive steps used for the computation of $\Psi_D$ and illustrated in Figure \ref{fig:FFTs_method_scheme}:\\
1) FFT of the array $C_0$. This leads to an array $D_1$ which contains $\Psi_D$ over a certain FoV of width $N_{D1}\,\lambda/D$ where $N_{D1}=N_{C0}$.\\ 
2) Selection of a central sub-array of $D_1$ to restrict the FoV to the area of interest of width $N_{D2}\,\lambda/D$. The dimensions of this new array $D_2$ are $N_{D2}\times N_{D2}$ with $N_{D2} \leq N_{D1}$. It contains $\Psi_D$ in a smaller FoV than in $D_1$.\\
3) Inverse FFT of the array $D_2$. It results in an array $C_2$ of same size as $D_2$ containing a miniature version of the pupil.\\
4) Use of the zero padding technique with the insertion of $C_2$ in a larger array $C_3$ of dimensions $N_{C3}\times N_{C3}$.\\
5) FFT of the array $C_3$. The result of this operation is an array $D_3$, of the same dimensions as $C_3$, containing the final complex amplitude $\Psi_D$ in plane D.\\
Thanks to these successive operations, a fast computation of $\Psi_D$ is achieved. To check the accuracy of our method, we compare the monochromatic intensity $|\Psi_D|^2$ obtained with our ultrafast Fourier transform method and with a simple FFT (steps 1 to 3 are skipped here). We work with $N_{C0}=390$ and $N_{D3}=10\,N_{C0}$ so the element resolution $\lambda/D$ is sampled each 10\,pixels in the image plane. This leads to an absolute difference less than $2\,\cdot 10^{-15}$. This numerical noise remains largely below all the contrast values that we are considering here and therefore, it does not have an influence on the analysis of our results. This test allows us to confirm the accuracy of our ultrafast Fourier transform proposed here. In the following, we apply this algorithm to estimate $\Psi_D$.

\subsection{Optimization criteria}
\begin{figure}[!ht]
\centering
\resizebox{\hsize}{!}{\includegraphics{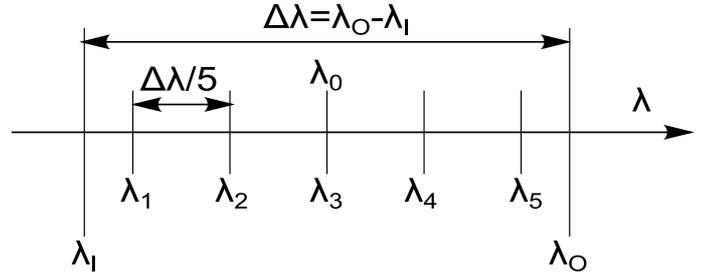}}
\caption{Schematic illustration of the wavelengths and bandwidth of study. $\lambda_I$ and $\lambda_O$ represent the limit wavelengths of the spectral bandwidth.} 
\label{fig:bandwidth_scheme}
\end{figure}

In their study, \citet{2003A&A...403..369S} optimized the parameters of the DZPM coronagraph to achieve the best on-axis extinction of the stellar diffraction pattern in the image plane. We decide here to optimize the CAPM coronagraph parameters in order to reach the best off-axis cancellation of the star image. This will yield minimal starlight pollution in the search area in which exoplanets are expected to be found. The search area is defined by an annulus of width between 2\,$\lambda_0$/D and 10\,$\lambda_0$/D from the main optical axis in the image plane.\\   
We average the intensity of the monochromatic coronagraphic image over this search area. This calculus is performed at five different wavelengths uniformly distributed within the spectral bandwidth. Once this done, we compute a merit function $M$ to estimate the quadratic sum of the mean intensities reached at each wavelength $\lambda$:
\begin{equation}
M=\sum_{\lambda_1}^{\lambda_n}{\sum_{\rho_I}^{\rho_O}{|\Psi_D|^2}}\,,
\end{equation}
where $\rho_I$ and $\rho_O$ are the inner and outer radii of the search area and $\lambda_{1}$ and $\lambda_{n}$ represent the first and n-th wavelengths of study within the spectral bandwidth. For our optimization, 5 wavelengths from $\lambda_1$ to $\lambda_5$ have been considered within the spectral bandwidth $\Delta\lambda$, centered on $\lambda_3=\lambda_0$, and equally spaced from one to another by $\Delta\lambda/5$, see Figure \ref{fig:bandwidth_scheme}. Some test optimizations have also been run with an eleven wavelengths gridding for 25\% and 50\% bands, showing no improvement of the coronagraph performance with respect to the five wavelength sampling.\\
A numerical least-squares method is used to obtain the CAPM coronagraph parameters ($d_1$, $d_2$, $\varphi_1$, $\varphi_2$, $\omega_{1, \lambda}$, $\omega_{2, \lambda}$, $\beta_{\lambda}$) that minimize our merit function and therefore, achieve the lowest averaged intensity over the search area for simultaneously all the wavelengths.

\subsection{Computation of the broadband image in plane D}
Once the parameters optimized, we can calculate the broadband image obtained with the CAPM coronagraph using Eq.(\ref{eq:intensity_D}). Since $\omega_{1, \lambda}$, $\omega_{2, \lambda}$ and $\beta_{\lambda}$ are estimated for five wavelengths, the monochromatic images at those wavelengths can be computed. To sum more than five frames for our final broadband image, we need to know the parameter values at other wavelengths. An interpolation of the values is realized for each apodization parameter. After several tests, we decide to use a second degree polynomial function to fit the parameter values, more representative of the behavior of actual materials than higher order functions. In the following, all our interpolations are performed with a second degree polynomial function. Once this defined, we can estimate the values of the parameter at a given wavelength and compute its frame. We then sum several monochromatic images with small spectral separation between two successive frames and obtain our broadband image in plane D.

\section{Performance over large spectral bandwidths}\label{sec:performance}  
In this section, we study the behavior of the CAPM coronagraph and its theoretical performance for different spectral bandwidths. This investigation is made for a clear circular aperture but it can be generalized to any kind of entrance pupil geometry. The current study will be particularly interesting in the context of some exoplanet imagers under consideration, such as 1-2\,m class off-axis space telescopes equipped with an internal coronagraph \citep[see e.g.,][]{2010SPIE.7731E..65G,2010SPIE.7731E..67T,2010SPIE.7731E..68G}. In addition, the results reached with our concept are compared to those previously achieved by \citet{2003A&A...403..369S} with the DZPM coronagraph. This allows us to underline the contrast gain obtained with the use of a colored apodization instead of a grey one.

\subsection{Colored apodization}
\begin{figure}[!ht]
\centering
\resizebox{\hsize}{!}{\includegraphics{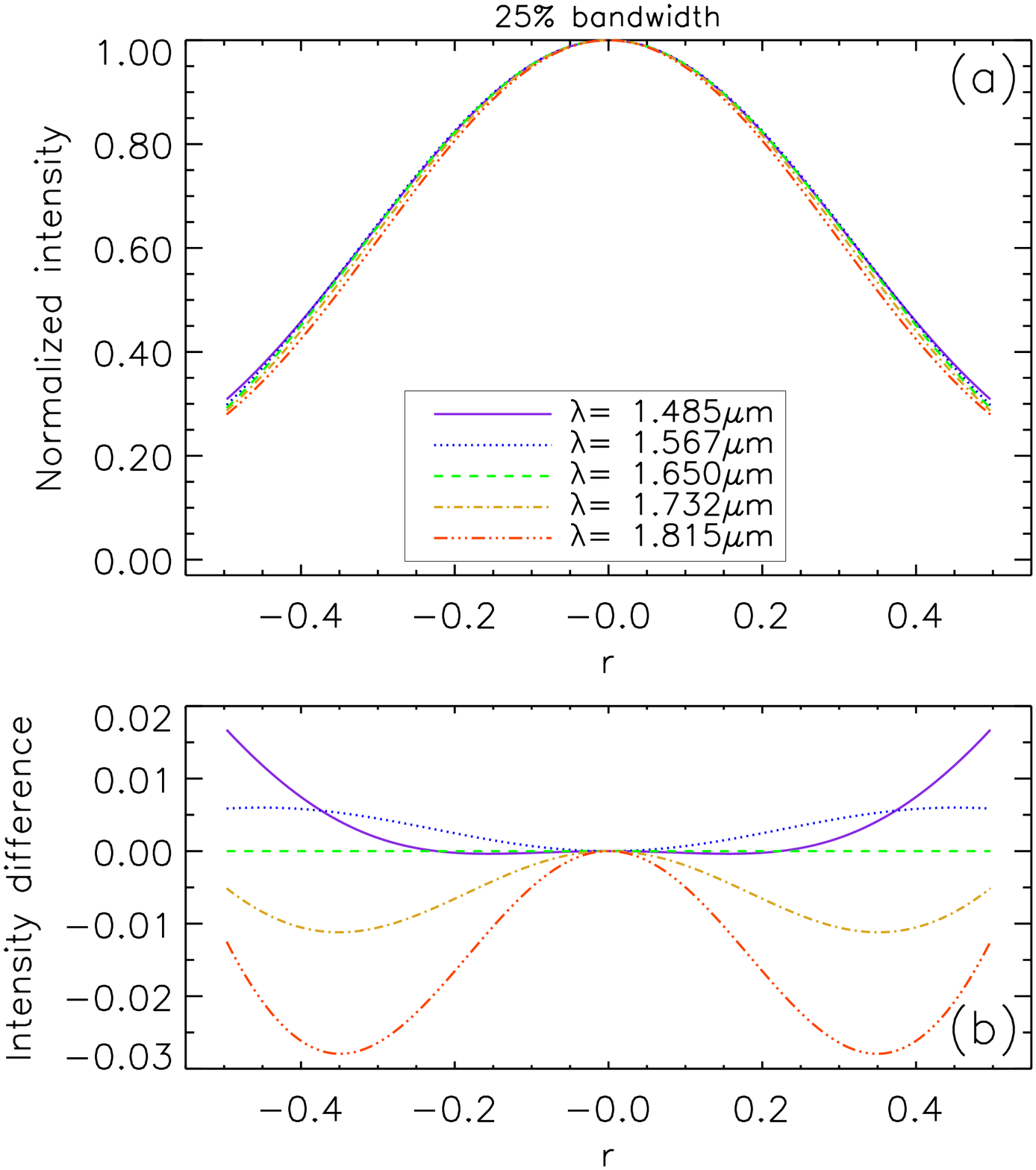}}
\resizebox{\hsize}{!}{\includegraphics{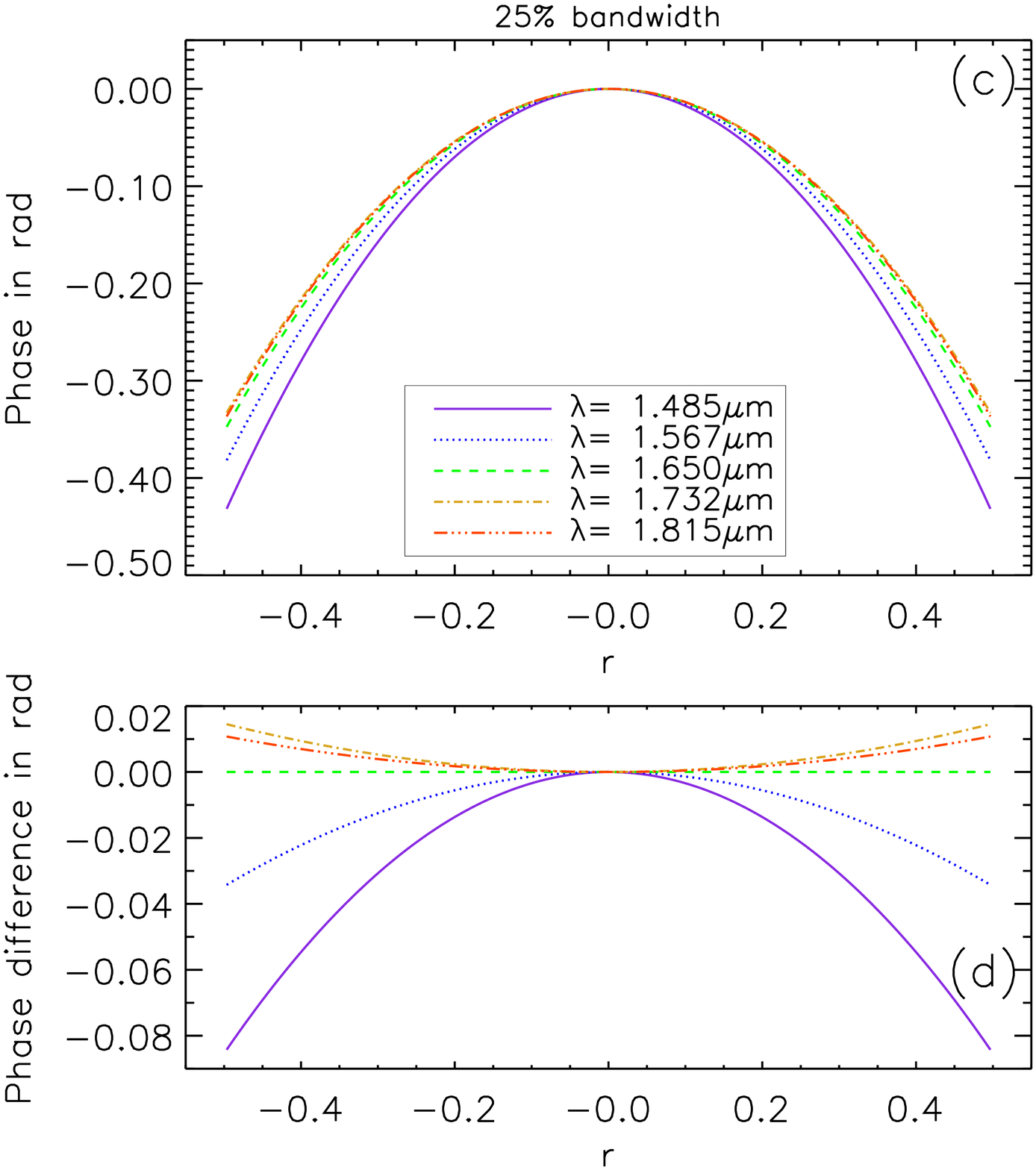}}
\caption{(a): radial profile of the colored intensity apodization for the CAPM coronagraph at different wavelengths $\lambda$. (b): radial profiles of the difference between the intensity apodization at a given wavelength $\lambda$ and that obtained at the central wavelength $\lambda_0$ (here $1.650\,\mu$m). (c) and (d) reproduce the panels (a) and (b) respectively, for the phase given by the phase apodization $\Phi_w$.} 
\label{fig:apod_example}
\end{figure}

Let us first observe the shape of the apodization reached for the CAPM coronagraph at different wavelengths. Figure \ref{fig:apod_example} panel (a) shows the radial intensity profiles of the colored apodization at five wavelengths for a 25\% bandwidth (H-band). We can notice the similarity between all the displayed profiles. To disentangle them, the difference between the transmission functions at a given wavelength $\lambda$ and at $\lambda_0$ has been drawn on the panel (b), emphasizing small difference in absolute value existing between different profiles. The intensity difference is not larger than 3\% of the normalized intensity peak but it reveals the chromatic aspect of our apodization used for the CAPM coronagraph. The apodization throughput evolves slightly from 57.9\% ($\lambda=1.485\,\mu$m) to 55.3\% ($\lambda=1.815\,\mu$m).\\
The panels (c) and (d) of Figure \ref{fig:apod_example} show the phase $\varphi$ of the phase apodization $\Phi_w=e^{i\varphi}$ in a similar way. The phase difference between the phases at a given wavelength $\lambda$ and at $\lambda_0$ is not larger than 0.1\,rad but it emphasizes the chromatic aspect of the phase apodization $\Phi_w$ in the CAPM coronagraph.

\subsection{Residual intensity}
\begin{figure}[!ht]
\centering
\resizebox{\hsize}{!}{\includegraphics{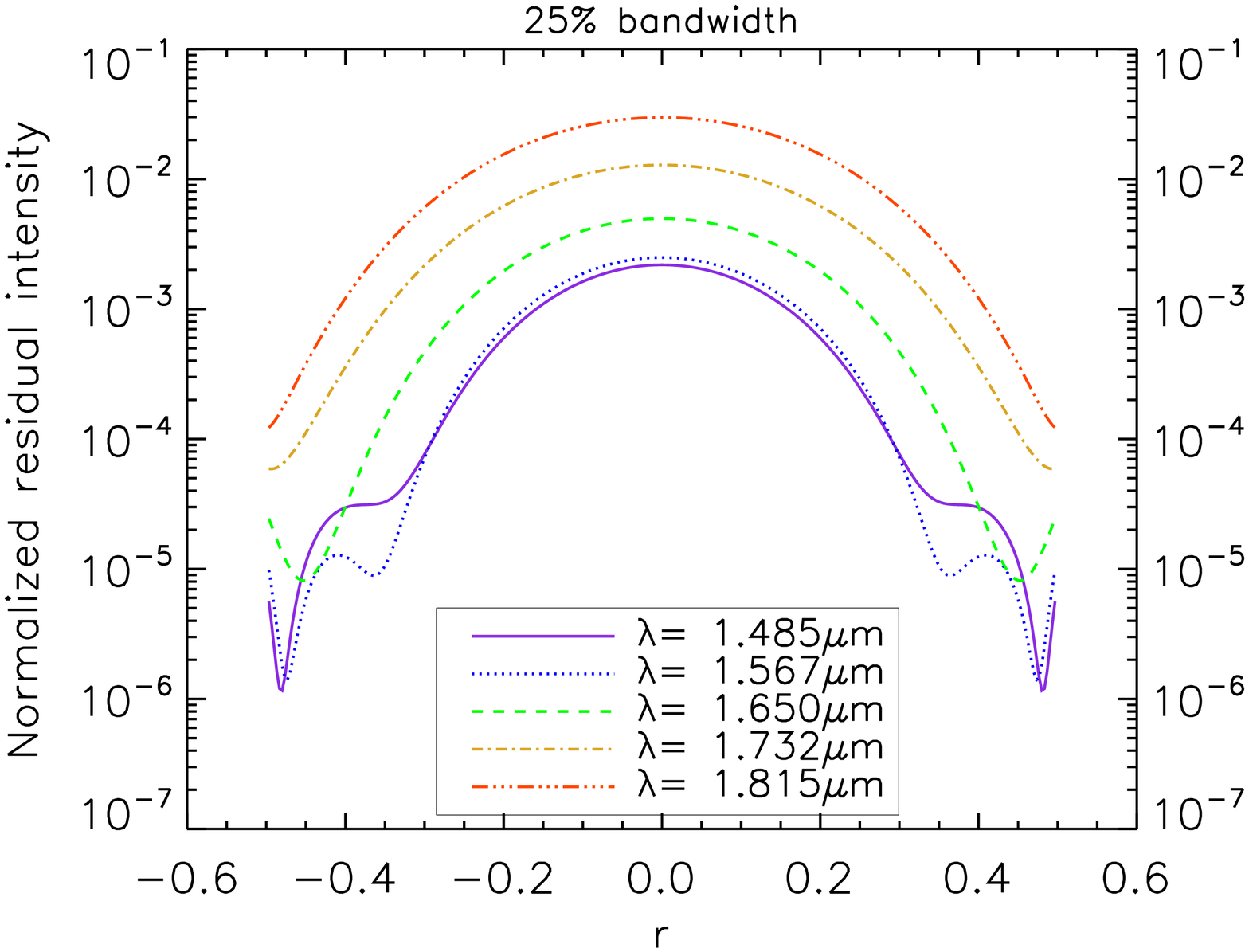}}
\resizebox{\hsize}{!}{\includegraphics{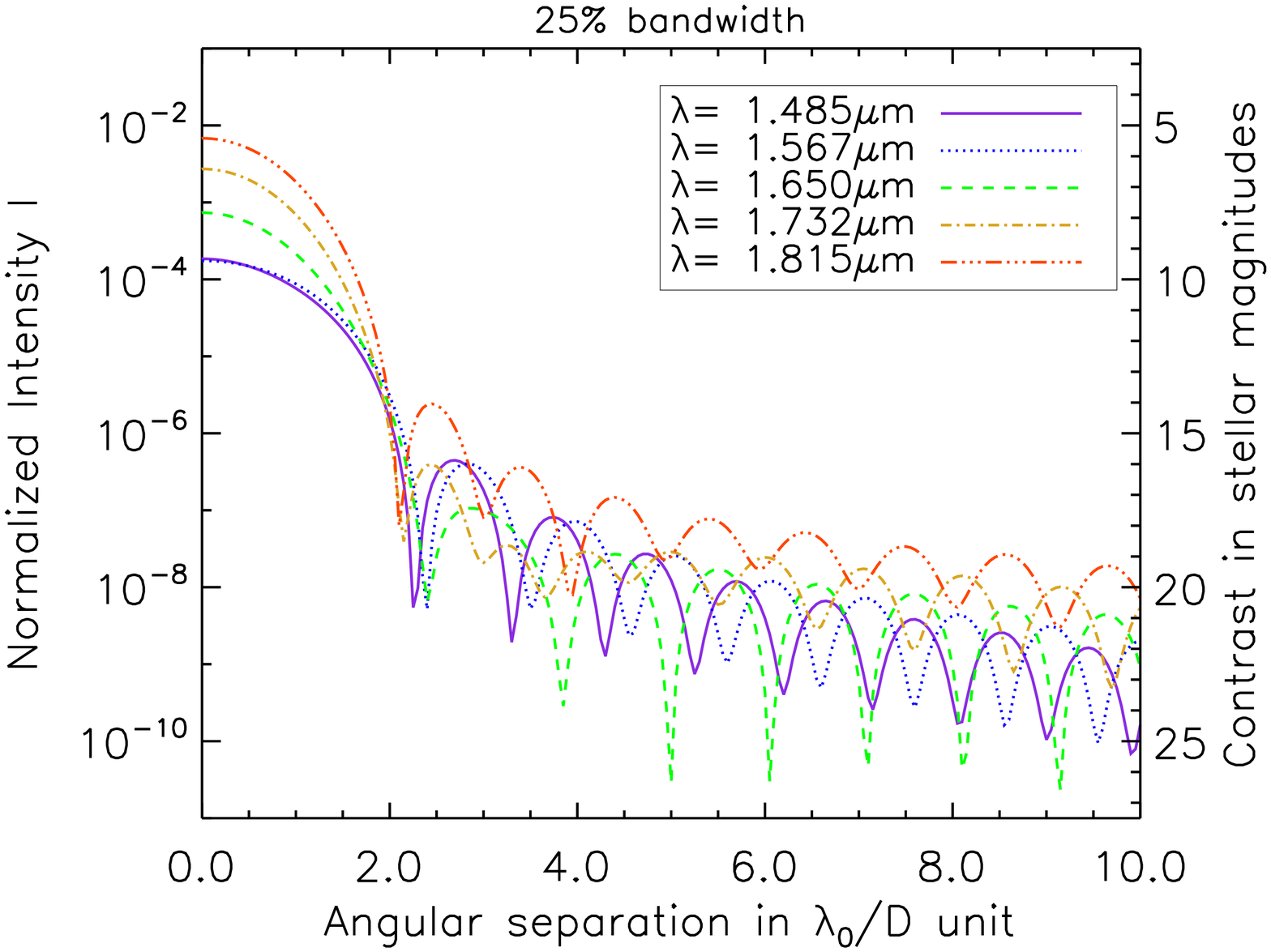}}
\caption{Radial profiles of the residual intensity with the CAPM coronagraph, within the re-imaged pupil in plane C (top) and in the final image plane D (bottom) at different wavelengths $\lambda$. Profiles have been normalized at the origin to directly read the coronagraphic extinction ratio at a given angular distance from the on-axis star image.} 
\label{fig:lyotstop_pupil_profiles}
\end{figure}

Much information can be extracted from the monochromatic intensity profiles in the Lyot and image planes, see Figure \ref{fig:lyotstop_pupil_profiles}. These curves have been obtained considering the optimized apodizer showed in Figure \ref{fig:apod_example}. The minimization of the residual broadband intensity inside the search area (2-10\,$\lambda_0/D$ range) has driven the optimization of our CAPM concept, leading to coronagraphic images with high intensity peak and low sidelobe level, see Figure \ref{fig:lyotstop_pupil_profiles} bottom plot. In parallel, the energy distribution inside the Lyot stop exhibits apodized intensity profiles, see Figure \ref{fig:lyotstop_pupil_profiles} top plot. Observing all the wavelengths, we note that the brightness of the sidelobes beyond 6\,$\lambda_0/D$ angular separation follows the intensity value found at the pupil edge. For sidelobes in the 2-6\,$\lambda_0/D$ range, this assertion is no longer valid because of the combination of several effects, including oscillations of the amplitude for some wavelengths in the Lyot plane as well as the presence of a phase apodization in the entrance pupil.
  
\subsection{Optimized parameters}
\begin{figure}[!ht]
\centering
\resizebox{\hsize}{!}{\includegraphics{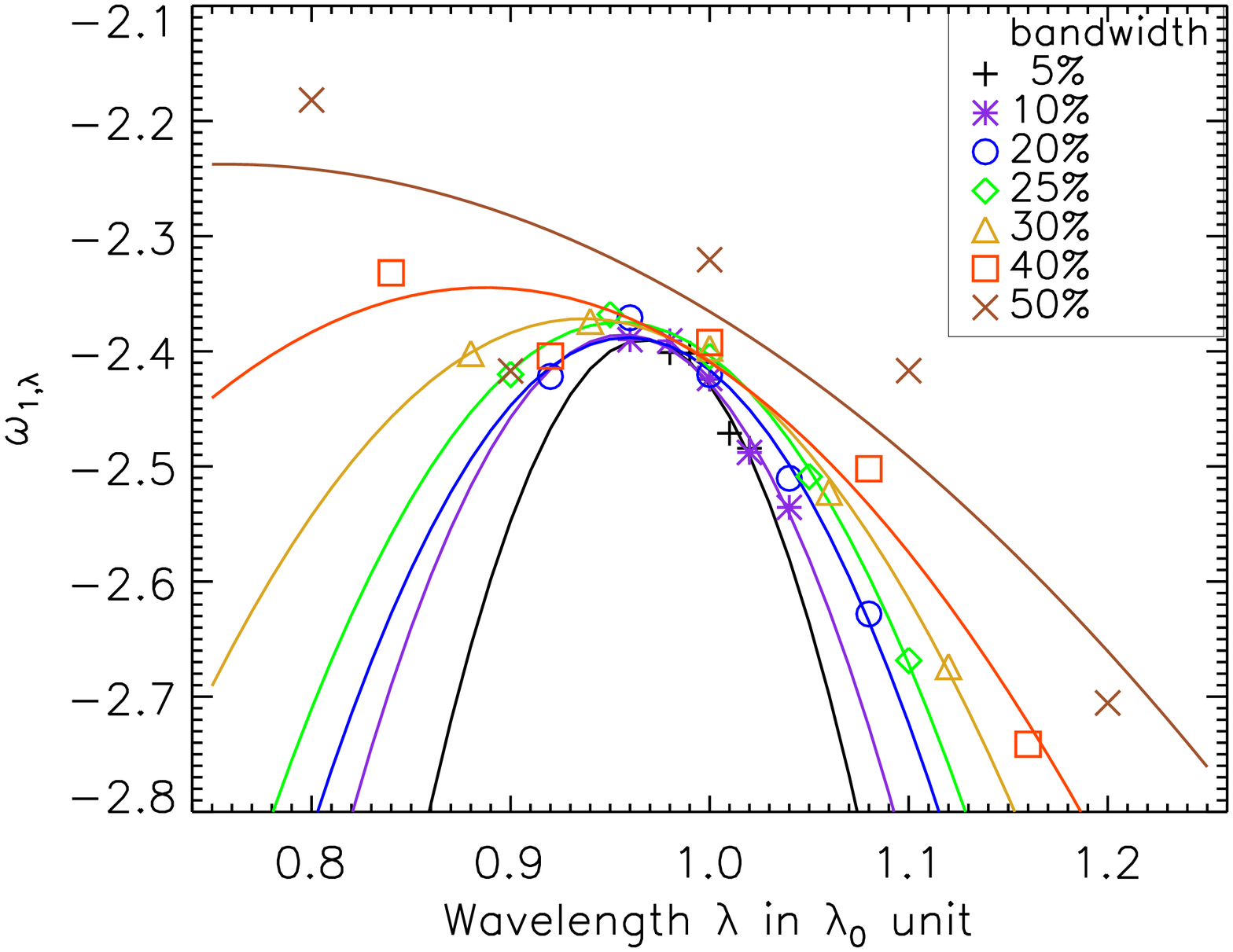}}
\resizebox{\hsize}{!}{\includegraphics{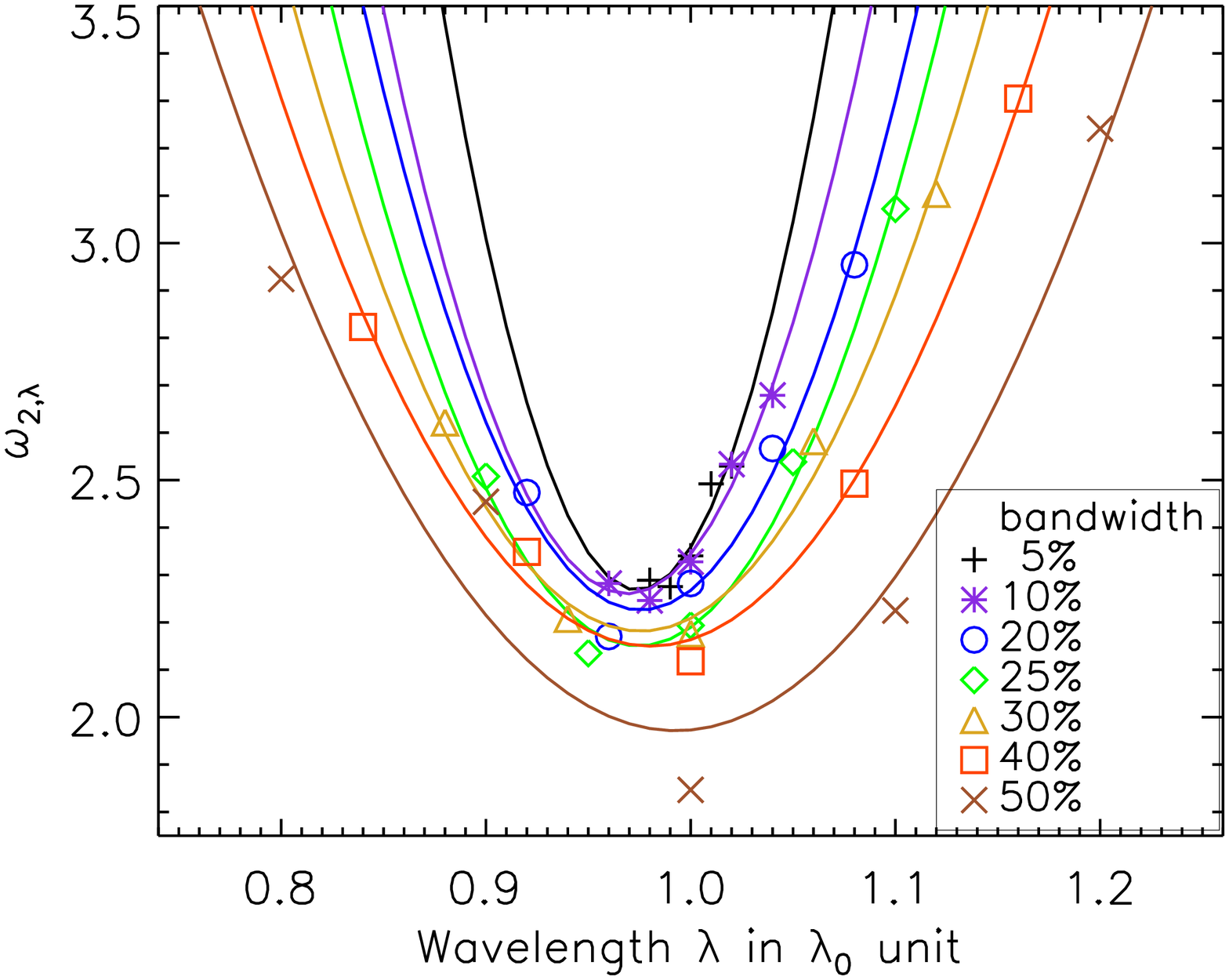}}
\resizebox{\hsize}{!}{\includegraphics{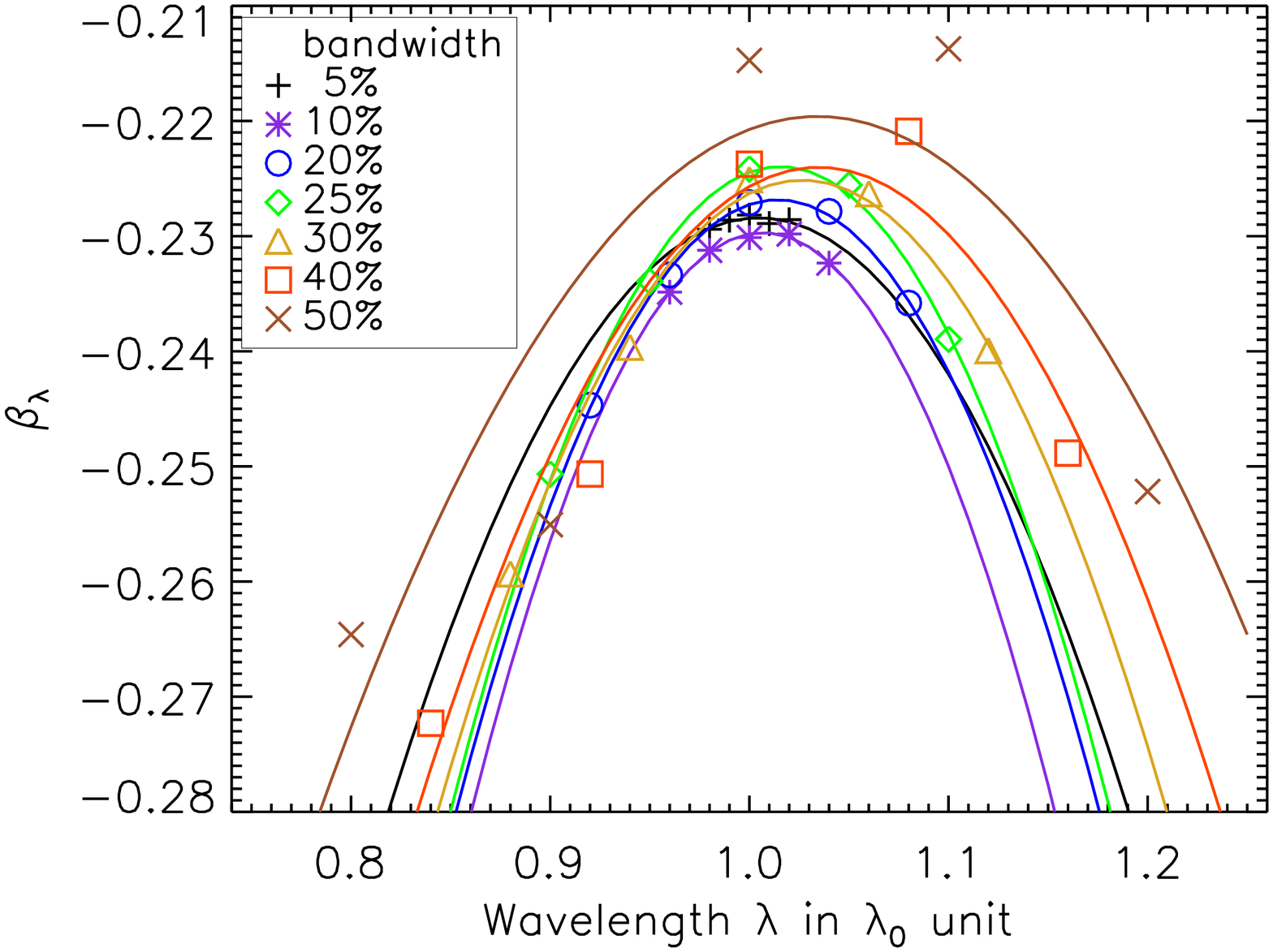}}
\caption{Optimized values of the amplitude apodization parameter $\omega_{1, \lambda}$ (top plot), $\omega_{2, \lambda}$ (middle) and the phase apodization parameter $\beta_{\lambda}$ (bottom) as a function of the wavelength for different bandwidths. A second-degree polynomial interpolation of the points is represented for each parameter and bandwidth.} 
\label{fig:omega12_beta_parameters}
\end{figure}

Following our optimization criteria, we estimate the different parameters of the CAPM coronagraph for different spectral bandwidths. The obtained values are given in Table \ref{table:parameter_values}. In addition, the colored apodization parameters $\omega_{1, \lambda}$, $\omega_{2, \lambda}$ and $\beta_{\lambda}$ are plotted as a function of the wavelength for different bandwidths in Figure \ref{fig:omega12_beta_parameters}.

\subsection{Contrast results and discussion}
\begin{figure}[!ht]
\centering
\resizebox{\hsize}{!}{\includegraphics{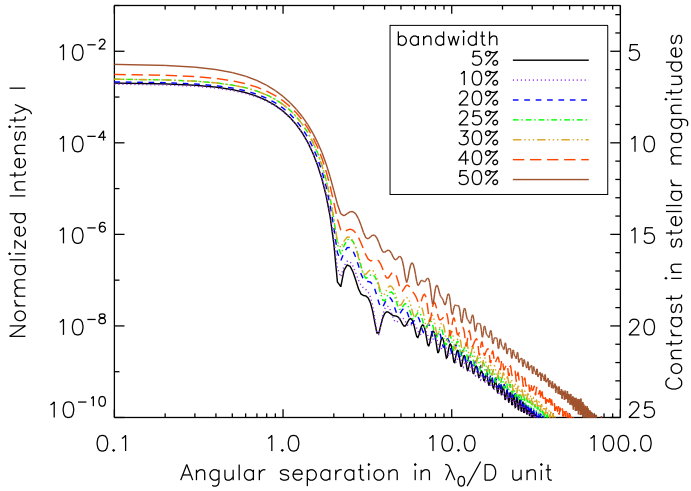}}
\caption{Radial intensity profile of the coronagraphic images achieved with the CAPM coronagraph for different sizes of the spectral bandwidth.} 
\label{fig:bandwidth_intensity_profiles}
\end{figure}

\begin{figure}[!ht]
\centering
\resizebox{\hsize}{!}{\includegraphics{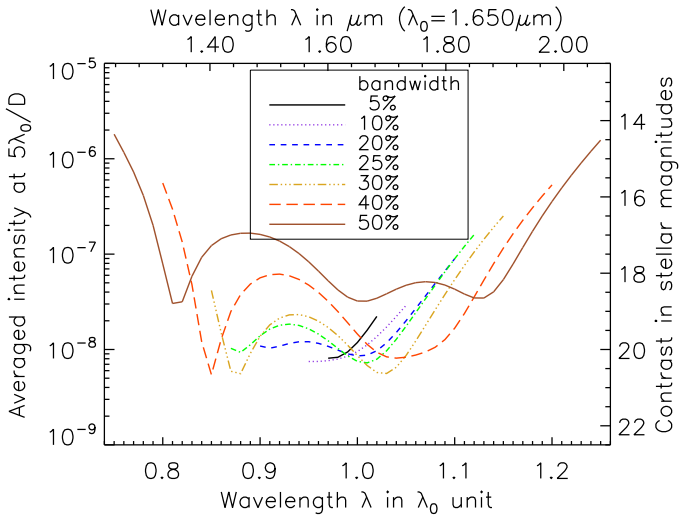}}
\caption{Theoretical averaged intensity at 5\,$\lambda_0/D$ obtained with the CAPM coronagraph as a function of the wavelength for different bandwidths.} 
\label{fig:bandwidth_averaged_intensity}
\end{figure}

\begin{figure}[!ht]
\centering
\resizebox{\hsize}{!}{\includegraphics{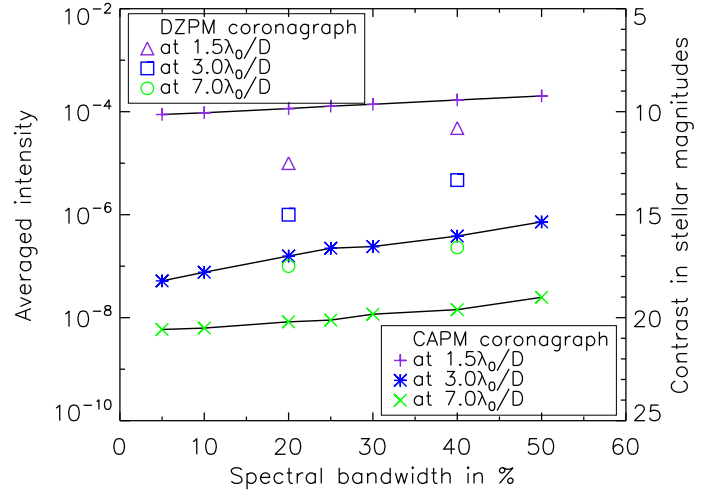}}
\caption{Theoretical averaged intensity achieved at different angular separations from the main optical axis with the DZPM and CAPM coronagraphs as a function of the bandwidth.} 
\label{fig:bandwidth_comparison}
\end{figure}

In Figure \ref{fig:bandwidth_intensity_profiles}, we draw the radial intensity profile of the images achieved with the CAPM coronagraph for different spectral bandwidths. The plotted curves are extracted from polychromatic images computed with N=1+$100\,\Delta\lambda/\lambda_0$ monochromatic frames with a separation corresponding to 1\% band. The values of the averaged intensities reached for each band and at several angular separations are given in Table \ref{table:parameter_values}. Averaged intensities of $2.2 \cdot 10^{-7}$ and $2.2 \cdot 10^{-8}$ are theoretically reached at 3$\,\lambda_0/D$ and 5$\,\lambda_0/D$ respectively, with our coronagraphic device for a 25\% bandwidth. As we can notice in Figure \ref{fig:omega12_beta_parameters}, the interpolation function is not optimal for the largest bands, leading to a loss of coronagraph performance when the optimized parameters are replaced by the interpolated values: the averaged intensity at $5\,\lambda_0/D$ degrades from $3.8\cdot 10^{-8}$ and $7.1\cdot 10^{-8}$ to $5.0\cdot 10^{-8}$ and $1.3\cdot 10^{-7}$ for the 40\% and 50\% bands respectively.\\
To go further in the analysis, we also determine the averaged intensity at $5\,\lambda_0/D$ achieved by the coronagraph at each wavelength of a given spectral bandwidth case, see Figure \ref{fig:bandwidth_averaged_intensity}. Each curve is plotted with the same 1\% sampling as described above. It can be noticed from this plot that a $2 \cdot 10^{-8}$ averaged intensity at $\sim\lambda_0$ can theoretically be achieved in the case of a 25\% bandwidth. Another interesting aspect about reachable averaged intensities concerns the spectral range. As the bandwidth decreases below 20\%, a single hole in the intensity curve appears around $\lambda_0$, surrounded by increasingly steep edges. In the monochromatic case, the coronagraph becomes identical to the ARPM coronagraph as we could expect.\\   
Finally in Figure \ref{fig:bandwidth_comparison}, we compare the results achieved with the CAPM coronagraph with those obtained by \citet{2003A&A...403..369S} with the DZPM coronagraph. It can be noticed that a $\sim$2.5\,$\,\magn$ contrast gain at 3\,$\lambda_0/D$ is reached with our device compared with the first DZPM coronagraph design for a 20\% and a 40\% bandwidths. Similar improvements of the reachable averaged intensities have been observed at other angular separations (5 and 7\,$\lambda_0/D$), confirming the substantial gain obtained for the CAPM coronagraph by replacing the grey by a colored apodization.  

\begin{table*}
\caption{Values of the CAPM coronagraph parameters optimized for different spectral bandwidths and averaged intensity results at different separations. The DZPM phase steps are given in optical path difference (OPD). Five wavelengths from $\lambda_1$ to $\lambda_5$ have been chosen within the spectral bandwidth $\Delta\lambda$, centered on $\lambda_3=\lambda_0$, and equally spaced from one to another by $\Delta\lambda/5$.}             
\label{table:parameter_values}      
\centering                          
\begin{tabular}{c c c c c c c c}        
\hline \hline
Parameters & \multicolumn{7}{c}{Bandwidth in \%}\\
 & 5 & 10 & 20 & 25 & 30 & 40 & 50\\
\hline
$d_1$ in $\lambda_0/D$ & 0.877 & 0.881 & 0.879 & 0.874 & 0.880 & 0.883 & 0.876\\
$d_2$ in $\lambda_0/D$ & 1.424 & 1.430 & 1.438 & 1.445 & 1.455 & 1.479 & 1.497\\
OPD$_1$ in $\lambda_0$ & 0.311 & 0.312 & 0.311 & 0.309 & 0.312 & 0.313 & 0.314\\
OPD$_2$ in $\lambda_0$ & 0.675 & 0.677 & 0.679 & 0.678 & 0.684 & 0.690 & 0.691\\
 & & & & & & & \\
$\omega_{1,\lambda}$ at $\lambda_1$ & -2.401 & -2.390 & -2.422 & -2.420 & -2.402 & -2.332 & -2.182\\
$\omega_{1,\lambda}$ at $\lambda_2$ & -2.402 & -2.391 & -2.371 & -2.368 & -2.374 & -2.404 & -2.417\\
$\omega_{1,\lambda}$ at $\lambda_3$ & -2.424 & -2.424 & -2.420 & -2.404 & -2.397 & -2.392 & -2.320\\
$\omega_{1,\lambda}$ at $\lambda_4$ & -2.471 & -2.488 & -2.510 & -2.509 & -2.522 & -2.502 & -2.417\\
$\omega_{1,\lambda}$ at $\lambda_5$ & -2.484 & -2.536 & -2.628 & -2.668 &  -2.674 & -2.741 & -2.705\\
 & & & & & & & \\
$\omega_{2,\lambda}$ at $\lambda_1$ & 2.289 & 2.282 &  2.474 & 2.508 &  2.621 & 2.823 & 2.924\\
$\omega_{2,\lambda}$ at $\lambda_2$ & 2.275 & 2.246 & 2.170 & 2.135 &  2.207 & 2.349 & 2.454\\
$\omega_{2,\lambda}$ at $\lambda_3$ & 2.340 & 2.328 & 2.282 & 2.193 & 2.176 & 2.118 & 1.847\\
$\omega_{2,\lambda}$ at $\lambda_4$ & 2.492 & 2.534 & 2.567 & 2.538 & 2.582 & 2.493 & 2.225\\
$\omega_{2,\lambda}$ at $\lambda_5$ & 2.530 & 2.679 & 2.954 & 3.072 & 3.104 & 3.305 & 3.241\\
 & & & & & & & \\
$\beta_{\lambda}$ at $\lambda_1$ & -0.229 & -0.235 & -0.245 & -0.251 & -0.259 & -0.272 & -0.265\\
$\beta_{\lambda}$ at $\lambda_2$ & -0.229 & -0.231 & -0.233 & -0.234 & -0.240 & -0.251 & -0.255\\
$\beta_{\lambda}$ at $\lambda_3$ & -0.228 & -0.230 & -0.227 & -0.224 & -0.225 & -0.224 & -0.215\\
$\beta_{\lambda}$ at $\lambda_4$ & -0.229 & -0.230 & -0.228 & -0.226 & -0.226 & -0.221 & -0.214\\
$\beta_{\lambda}$ at $\lambda_5$ & -0.229 & -0.232 & -0.236 & -0.239 & -0.240 & -0.249 & -0.252\\
 & & & & & & & \\
level at $1.5\,\lambda_0/D$ & 8.8$ \cdot 10^{-5}$ & 9.5$ \cdot 10^{-5}$ & 1.1$ \cdot 10^{-4}$ & 1.3$ \cdot 10^{-4}$ & 1.4$ \cdot 10^{-4}$ & 1.7$ \cdot 10^{-4}$ & 2.0$ \cdot 10^{-4}$\\
level at $2.0\,\lambda_0/D$ & 6.6$ \cdot 10^{-6}$ & 7.4$ \cdot 10^{-6}$ & 9.9$ \cdot 10^{-6}$ & 1.2$ \cdot 10^{-5}$ & 1.3$ \cdot 10^{-5}$ & 1.8$ \cdot 10^{-5}$ & 2.4$ \cdot 10^{-5}$\\
level at $3.0\,\lambda_0/D$ & 5.2$ \cdot 10^{-8}$ & 7.6$ \cdot 10^{-8}$ & 1.6$ \cdot 10^{-7}$ & 2.2$ \cdot 10^{-7}$ & 2.4$ \cdot 10^{-7}$ & 3.8$ \cdot 10^{-7}$ & 7.2$ \cdot 10^{-7}$\\
level at $5.0\,\lambda_0/D$ & 1.1$ \cdot 10^{-8}$ & 1.2$ \cdot 10^{-8}$ & 1.9$ \cdot 10^{-8}$ & 2.2$ \cdot 10^{-8}$ & 2.9$ \cdot 10^{-8}$ & 3.8$ \cdot 10^{-8}$ & 7.1$ \cdot 10^{-8}$\\
level at $7.0\,\lambda_0/D$ & 5.9$ \cdot 10^{-9}$ & 6.3$ \cdot 10^{-9}$ & 8.3$ \cdot 10^{-9}$ & 9.0$ \cdot 10^{-9}$ & 1.2$ \cdot 10^{-8}$ & 1.4$ \cdot 10^{-8}$ & 2.5$ \cdot 10^{-8}$\\
\hline
\end{tabular}
\end{table*}
%

\section{Design proposal for the colored apodization}\label{sec:design}
The realization of a colored apodizer is a key point for the implementation of the CAPM coronagraph. This requires careful control of the transmission function of the apodizer both spatially across the pupil and spectrally over a wide range of wavelengths. We expose here some of the options currently investigated for the fabrication of this chromatic apodization.\\
One option consists of using an absorbing material with an appropriate chromatic variation of the absorption coefficient, for instance a colored-glass filter. The required transmission function for the apodizer is obtained at all the wavelengths by careful thickness control of the absorbing material. Appropriate material choice would lead to a negative lens of up to a few mm edge thickness. The optical power of this lens could possibly be included in the optical design of the instrument, but in a more general option it would need to be compensated by a positively powered non-absorbing lens.\\ 
Another approach for the colored apodizer that we have considered is to use a thin film of a highly absorbing material such as metal, deposited with an appropriate radial thickness profile onto a silica substrate. The chromaticity of the apodizer is now related to the chromatic variation of the complex refractive index of the material.\\ 
Solutions based on metallic thin film, alloy layer or else High-Energy Beam Sensitive (HEBS) glass have already been investigated to produce achromatic band-limited masks in the visible \citep{2007ApOpt..46.7485S,2008SPIE.7010E.107M,2008ApOpt..47..116B} or manufacture achromatic apodizers of the APLC coronagraph for VLT-SPHERE \citep{2008SPIE.7014E.116G,2011ExA....30...59G,2011ExA....30...39C} and Gemini Planet Imager \citep{2009SPIE.7440E..23S,2009SPIE.7440E..40S}. Our study is focused here on the search of metals to design chromatic components instead of grey masks.\\  
In the following, we detail this approach for the realization of the colored apodization. The performance of the CAPM coronagraph is estimated with this physical design and then compared to the results with the theoretical model of colored apodizer described above. We assume a clear circular aperture and a 25\% bandwidth centered at $\lambda_0=1.65\,\mu$m (H-band) for our optimization.

\subsection{Assembly based on a metallic thin film on a silica substrate}
\begin{figure}[!ht]
\centering
\resizebox{\hsize}{!}{\includegraphics{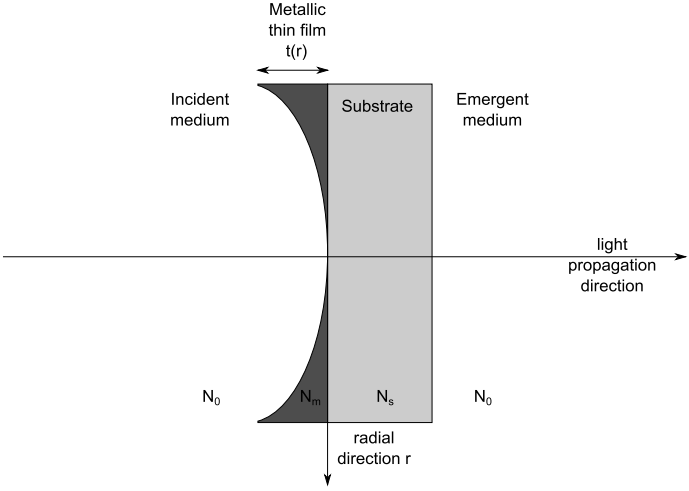}}
\caption{Schematic illustration of the assembly, composed of a metallic thin film deposited on a substrate. The scheme is not at scale.} 
\label{fig:assembly_scheme}
\end{figure}

For the manufacturing of our colored amplitude apodization $\Phi_a$, we investigate an assembly based on a metallic thin film deposited on a silica substrate with complex refractive index $N_m$ and $N_s$ respectively, see Figure \ref{fig:assembly_scheme}. At the boundaries of our assembly, we consider an absorption-free medium of refractive index $N_0$. The assembly is assumed to work in transmission at normal incidence.\\
The substrate is supposed to be an absorption-free medium, thick enough to neglect the internal reflections. Referring to the metallic thin film, the refractive index $N_m$ presents a complex part, called extinction coefficient $k$, which is related to the absorption $\alpha_m$ within the metal layer by $\alpha_m=4\pi k/\lambda$. Increasing the metal thickness leads to a combined enhancement of the absorption and reflection of the assembly and therefore, a reduction of its transmission. A careful control of the metallic thin film thickness allows us to adjust the amplitude transmission function of our assembly. The chromaticity of the amplitude apodization is mainly due to the dependence on wavelength of the metal extinction coefficient.\\ 
For the present concept, the amplitude transmission function of our assembly $\Phi_{b}$ is defined as the product of the transmittances through the metallic thin film and the substrate $\tau_m$ and $\tau_s$ respectively. The amplitude apodization can therefore be written as:
\begin{equation}
\begin{split}
\Phi_{b}(r, \lambda)= & \tau_m(r, \lambda) \tau_s(\lambda) \,,\\
\end{split}
\label{eq:model_2}
\end{equation}
For the sake of clarity, we omit the dependence on wavelength of the refractive indices and the parameters $B$, $C$ and $\delta_m$ which appear in the following equations. The transmittances in amplitude $\tau_m$ and $\tau_s$ can be expressed as: 
\begin{equation}
\begin{split}
\tau_m(r, \lambda)= & 2N_0/(N_0 B+C)\,,\\
\tau_s(\lambda)= & 2N_s/(N_s+N_0)\,,
\end{split}
\label{eq:transmittance_5}
\end{equation}
where $B$ and $C$ can be identified as the normalized electric and magnetic field amplitudes at the front interface respectively \citep{1991tfof.book.....M}. In our case, $B$ and $C$ can be calculated with the following equation:
\begin{equation}
\begin{pmatrix}
B\\C 
\end{pmatrix}
=
\begin{pmatrix}
\cos\delta_m & (i\sin\delta_m)/N_m\\
i N_m\sin\delta_m & \cos\delta_m 
\end{pmatrix}
\begin{pmatrix}
1\\N_s
\end{pmatrix}
\label{eq:BC_5}
\end{equation}
with $\delta_m(r, \lambda)= 2\pi N_m t(r)/\lambda$ the phase shift experienced by the wave as it traverses a distance $t$ normal to the substrate boundary.\\
The complex apodization $\Phi$ of the CAPM coronagraph with the physical design writes as: 
\begin{equation}
\Phi=\Phi_b\cdot\Phi_w \,.
\label{eq:decomp_apodization_assembly}
\end{equation}
The metallic thin film also induces a radially variable phase delay, leading to complex values evolving with $r$ for the transmission function $\Phi_b$. Since our concept of CAPM coronagraph also contains a phase apodization $\Phi_w$, the total phase of the complex apodization $\Phi$ in the entrance pupil plane A will be the combination of the phase terms of $\Phi_w$ and $\Phi_b$.

\subsection{Choice of the metal}
\begin{figure}[!ht]
\centering
\resizebox{\hsize}{!}{\includegraphics{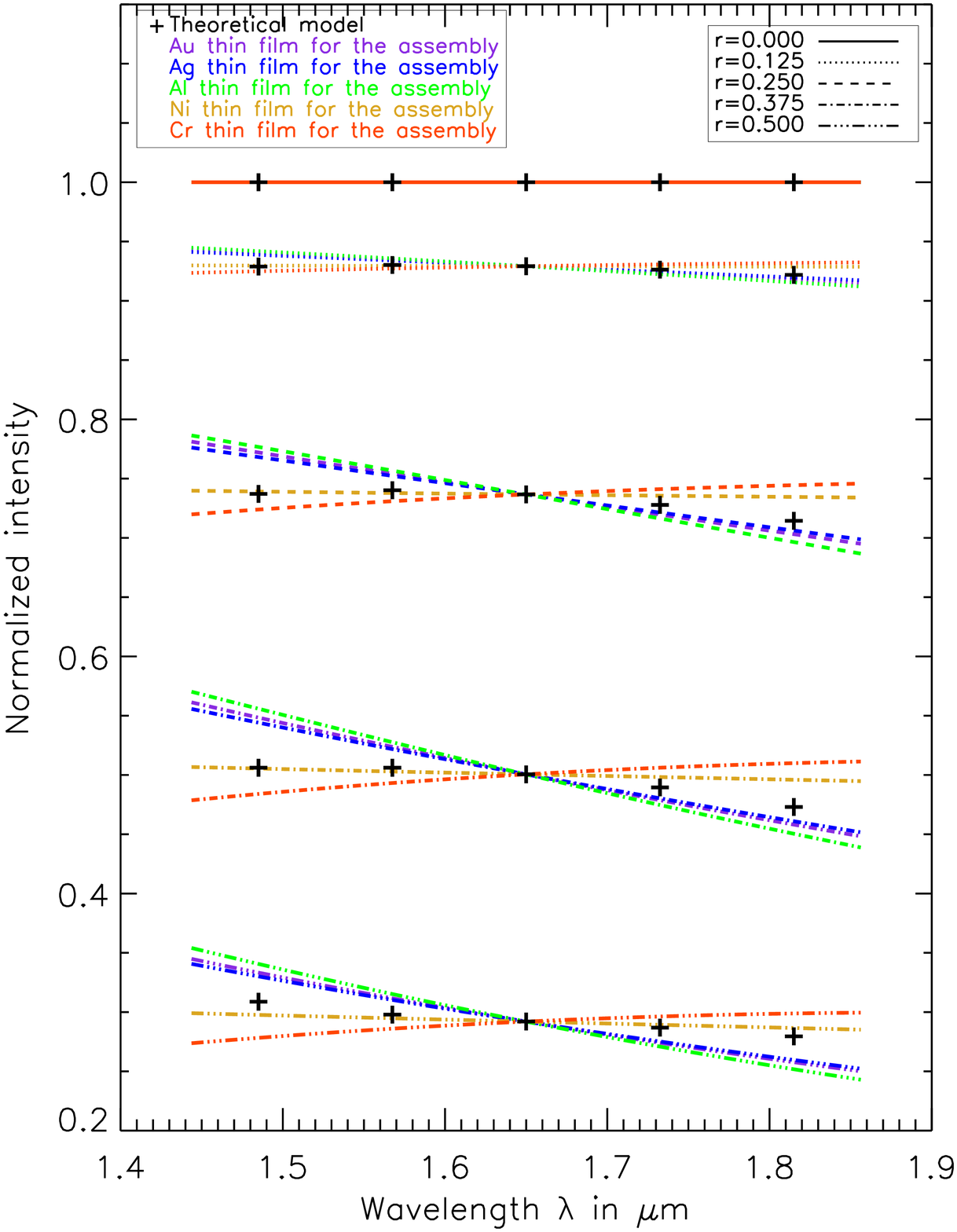}}
\caption{Transmission values in intensity obtained for the theoretical model and the assembly with different metals as a function of the wavelength. Values are provided for different normal radial distances. Thickness for each material and radial distance are given in Table \ref{table:thickness_values_for_models}.} 
\label{fig:metal_profile}
\end{figure}

Preliminary studies have been realized to choose the metal for the physical model, see Figure \ref{fig:metal_profile}. Five metals (gold, silver, aluminum, nickel and chromium) have been considered for this purpose while five radial distances have been investigated to determine the optimal thickness of the metallic thin film at different positions in the pupil. For each metal and radial distance, the thickness of the metallic thin film has been estimated such that the assembly reaches the same intensity transmission as the theoretical model at $\lambda_0$ observed in Figure \ref{fig:apod_example}. Bulk refractive indices for the metal layer, obtained from \citet{palik1985handbook}, have been used for the implementation of our simulations. The thicknesses reached for each metal and radial distance are given in Table \ref{table:thickness_values_for_models}.\\
Using silver or nickel provides the best fit to the theoretical model at small or large radial distances, respectively. However, even with those metals, the theoretical profile of the colored apodization is not reached at all the wavelengths of study. In the following, we re-optimize the CAPM coronagraph, using a nickel apodizer.\\
With reference to the apodizer model of Eq. (\ref{eq:ampl_apodization}), the thickness profile $t$ of the metal is optimized using the following expression:
\begin{equation}
t(r)= t_c \left [\chi_{1}\left (r^2-\frac{\eta^2}{4}\right ) + \chi_{2}\left (r^4-\frac{\eta^4}{16}\right ) \right ] \,,
\label{eq:thickness_2}
\end{equation}
where $t_c=1$\,nm is the reference thin film thickness and $\chi_{1}$ and $\chi_{2}$ are the parameters related to the terms in $r^2$ and $r^4$ respectively of the thickness profile. This assembly provides a maximum amplitude transmission at the center of the pupil or at the edge of the central obscuration for $\eta \neq 0$.\\

\begin{table}[!ht]
\caption{Thickness of the different metals for the assembly at different radial distances used in Figure \ref{fig:metal_profile}.}             
\label{table:thickness_values_for_models}      
\centering                          
\begin{tabular}{c r r r r r}        
\hline \hline
\multirow{2}*{Radial distance} & \multicolumn{5}{c}{Metallic thin film thickness in nm}\\
 & Au & Ag & Al & Ni & Cr\\
\hline
$r=0.125$ & 1.044 & 1.237 & 0.324 & 0.485 & 0.557\\
$r=0.250$ & 2.822 & 3.228 & 0.977 & 2.070 & 2.449\\
$r=0.375$ & 5.081 & 5.733 & 1.844 & 4.952 & 6.092\\
$r=0.500$ & 8.177 & 9.148 & 3.062 & 9.666 & 12.403\\
\hline
\end{tabular}
\end{table}
%

\subsection{Results}
\begin{figure}[!ht]
\centering
\resizebox{\hsize}{!}{\includegraphics{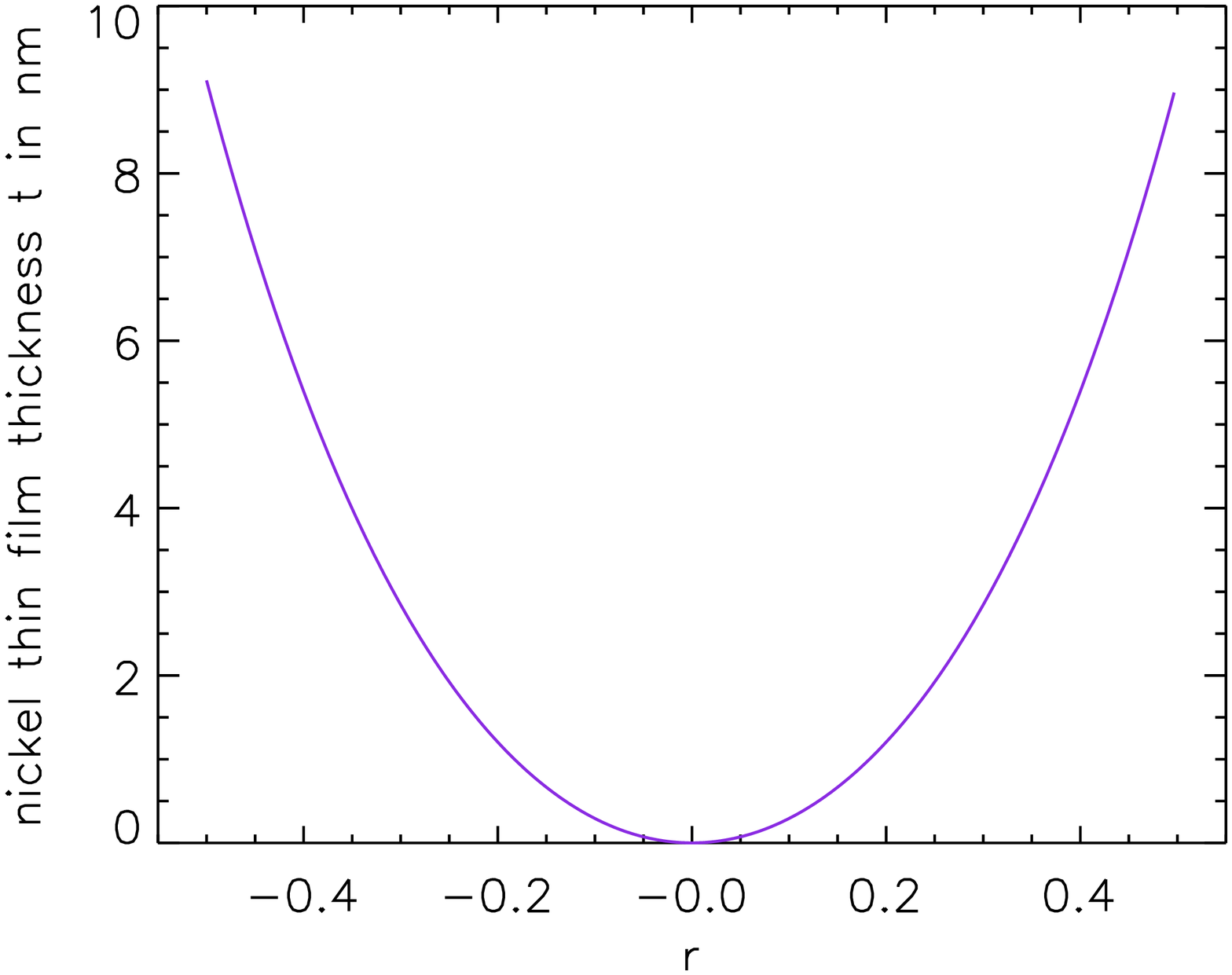}}
\resizebox{\hsize}{!}{\includegraphics{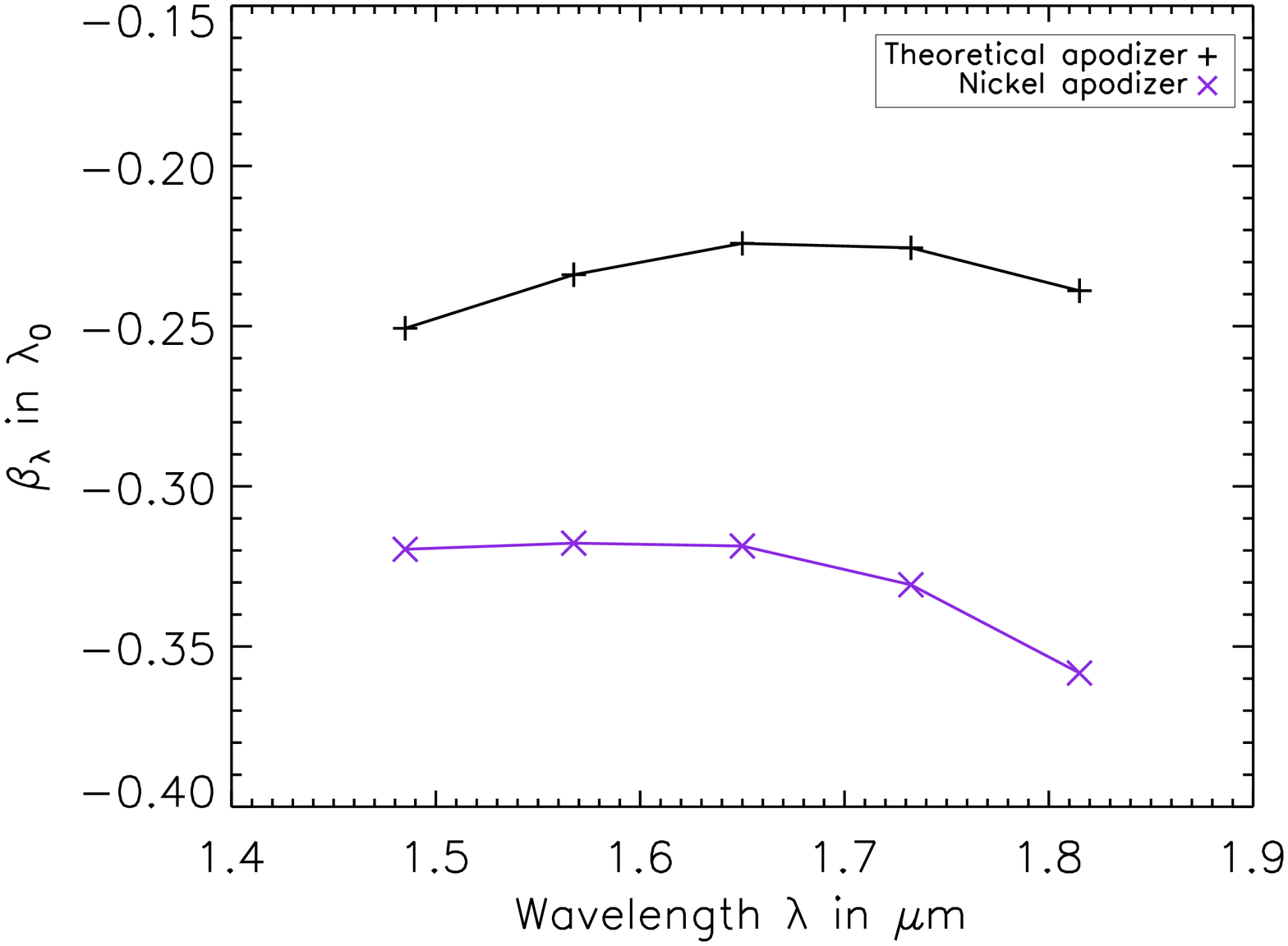}}
\caption{Top: thickness profile of the nickel layer as a function of the wavelength. Bottom: optimized values of the phase apodization parameter $\beta_{\lambda}$ for the CAPM coronagraph in the presence of the theoretical model and the physical design as a colored amplitude apodization.} 
\label{fig:thickness_beta_design_parameters}
\end{figure}

\begin{figure}[!ht]
\centering
\resizebox{\hsize}{!}{\includegraphics{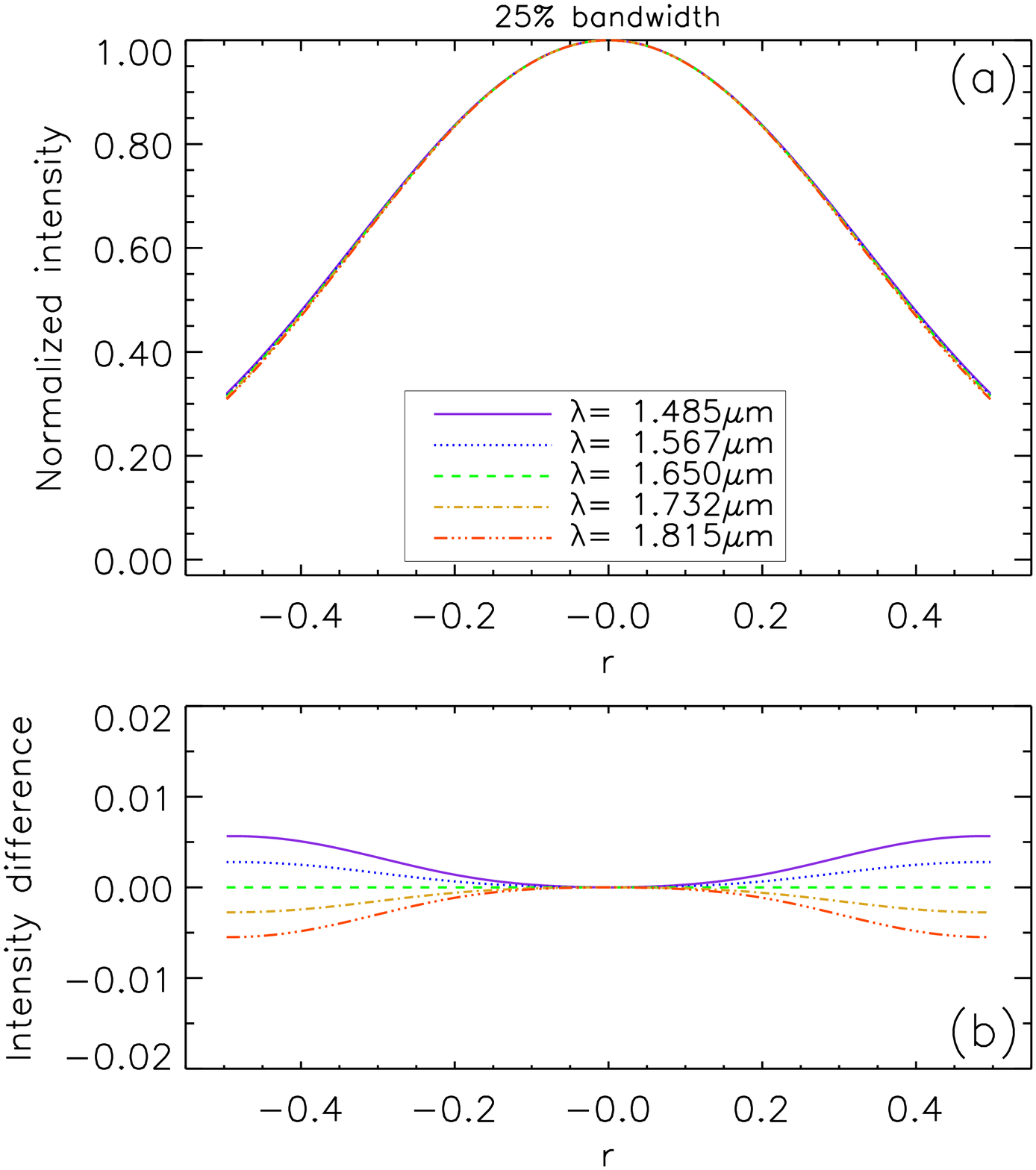}}
\resizebox{\hsize}{!}{\includegraphics{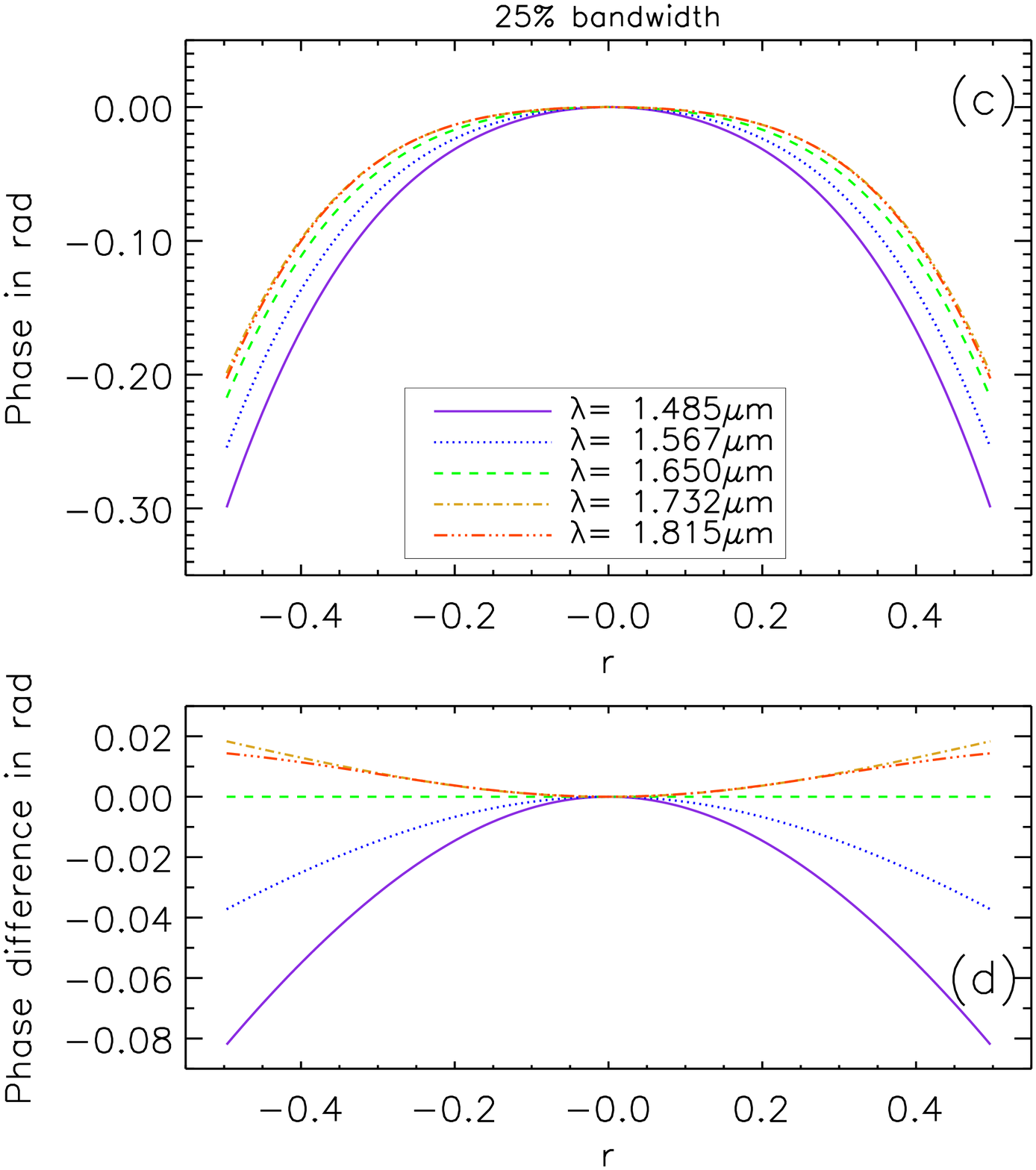}}
\caption{(a): radial intensity profile of the colored apodizer, made with an assembly using nickel layer, for the CAPM coronagraph at different wavelengths $\lambda$. (b): radial profiles of the difference between the intensity apodization at a given wavelength $\lambda$ and that obtained at the central wavelength $\lambda_0$ (here $1.650\,\mu$m). (c) and (d) reproduce (a) and (b) respectively for the phase given by the phase apodization $\Phi_w$.} 
\label{fig:apod_assembly_design}
\end{figure}

\begin{figure}[!ht]
\centering
\resizebox{\hsize}{!}{\includegraphics{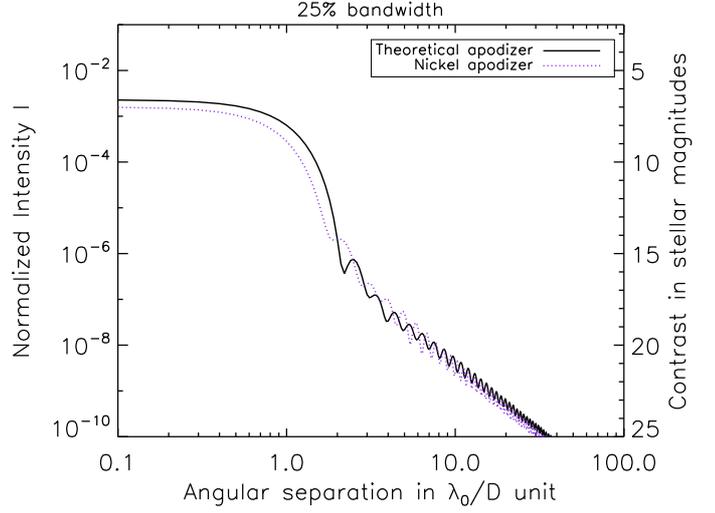}}
\caption{Radial intensity profile of the coronagraphic images achieved with the CAPM coronagraph using the theoretical model and the physical design for the colored apodizer.} 
\label{fig:Iprofile_assembly}
\end{figure}

The optimized parameters for the phase mask and apodizer are reported in Table \ref{table:parameter_values_for_models} while the resulting thickness profile $t$ of the nickel film and the parameter $\beta_\lambda$ associated to the phase apodization are plotted in Figure \ref{fig:thickness_beta_design_parameters}. Regarding $t$ displayed in the top plot, a maximum thickness of 9.1\,nm is reached at the pupil edge for the metal layer, proving to be consistent with the 9.7\,nm value observed at $r=0.5$ in the tests for the choice of the metal and reported in Table \ref{table:thickness_values_for_models}. Concerning the $\beta_\lambda$ curve in the bottom plot, we can note some discrepancy between the values obtained for the CAPM coronagraphs working with the theoretical model and the physical design. This difference can be explained by the presence of the additional phase shifting introduced by the metallic thin film of the assembly, leading to a re-adjustment of the parameter $\beta_\lambda$ of the chromatic defocus during the optimization of the CAPM coronagraph with the physical design.\\ 
The radial intensity profiles of the apodizer for a 25\% bandwidth (H-band) are represented at five wavelengths in Figure \ref{fig:apod_assembly_design} panel (a). While the profiles are quite similar to the profiles shown in Figure \ref{fig:apod_example}, the difference between them, shown in panel (b), are smaller than for the optimal profiles. The apodization throughput of the assembly ranges from 59.5\% ($\lambda=1.485\,\mu$m) to 58.8\% ($\lambda=1.815\,\mu$m), which is slightly higher than the theoretical colored apodizer. The phase profiles, shown in panel (c), are somewhat different from those of the optimal apodizer in Figure \ref{fig:apod_example}, indicating the presence of a fourth-order term (spherical aberration) in addition to the second order term. The differential phase profiles in panel (d) are remarcably similar to those of the optimal apodizer, both in shape and amplitude.\\
In Figure \ref{fig:Iprofile_assembly}, we represent the intensity profiles obtained with the CAPM coronagraph in the presence of the theoretical colored apodizer and our physical design. The profiles are almost identical, underlining the good relevance of using a nickel layer to realize our colored amplitude apodization. The intensity levels estimated at different separations from the star are thus very close for the two models of colored amplitude apodization, see Table \ref{table:contrast_values_for_models}. For instance, the intensity levels at 5\,$\lambda_0/D$ are $2.6\cdot 10^{-8}$ and $2.2\cdot 10^{-8}$, respectively. An optimized design using a nickel thin film therefore allows us to reproduce the contrast results obtained for the optimal CAPM coronagraph. The nickel layer presents a maximum thickness of 9.1\,nm with an accurate control at atomic level. To manufacture such a design, atomic layer deposition (ALD) represents a very attractive technique, widely used in microelectronics, since it enables the production of very thin, conformal film with thickness control and composition of the films at the atomic level for a large range of materials \citep{Leskelä2002138}.\\
Further studies and tests will be required for the manufacturing of such chromatic apodizers. Finally, options based on more elaborate, radially variable thin film stacks, possibly containing both dielectric and metallic materials, are also being considered.

\begin{table}[!ht]
\caption{Values of the parameters for the CAPM coronagraph optimized with an assembly using nickel thin film for the colored apodization. The DZPM phase steps are given in optical path difference (OPD). Five wavelengths from $\lambda_1$ to $\lambda_5$ have been chosen within the spectral bandwidth $\Delta\lambda$, centered on $\lambda_3=\lambda_0$, and equally spaced from one to another by $\Delta\lambda/5$.}             
\label{table:parameter_values_for_models}      
\centering                          
\begin{tabular}{c c}        
\hline \hline
Parameters & 25\% bandwidth\\
\hline
$d_1$ in $\lambda_0/D$ & 0.868\\
$d_2$ in $\lambda_0/D$ & 1.478\\
OPD$_1$ in $\lambda_0$ & 0.338\\
OPD$_2$ in $\lambda_0$ & 0.716\\
 & \\
$\chi_1$ & 28.9\\
$\chi_2$ & 30.2\\
 & \\
$\beta_{\lambda}$ at $\lambda_1$ & -0.320\\
$\beta_{\lambda}$ at $\lambda_2$ & -0.318\\
$\beta_{\lambda}$ at $\lambda_3$ & -0.319\\
$\beta_{\lambda}$ at $\lambda_4$ & -0.331\\
$\beta_{\lambda}$ at $\lambda_5$ & -0.358\\ 
\hline
\end{tabular}
\end{table}
%

\begin{table}[!ht]
\caption{Intensity values provided by the CAPM coronagraph with the theoretical model and the physical design for the colored apodization. Results are given for a 25\% bandwidth centered at $\lambda_0=1.650\,\mu$m (H-band).}             
\label{table:contrast_values_for_models}      
\centering                          
\begin{tabular}{c c c}        
\hline \hline
\multirow{2}*{level at} & \multicolumn{2}{c}{CAPM coronagraph with}\\
 & theoretical apodizer & assembly\\
\hline
1.5\,$\lambda_0/D$ & 1.3$ \cdot 10^{-4}$ & 1.3$ \cdot 10^{-4}$\\
2.0\,$\lambda_0/D$ & 1.2$ \cdot 10^{-5}$ & 1.2$ \cdot 10^{-5}$\\
3.0\,$\lambda_0/D$ & 2.2$ \cdot 10^{-7}$ & 1.4$ \cdot 10^{-7}$\\
5.0\,$\lambda_0/D$ & 2.2$ \cdot 10^{-8}$ & 1.9$ \cdot 10^{-8}$\\
7.0\,$\lambda_0/D$ & 9.0$ \cdot 10^{-9}$ & 8.2$ \cdot 10^{-9}$\\ 
\hline
\end{tabular}
\end{table}
%

\subsection{Implementation of the chromatic defocus}
\begin{figure}[!ht]
\centering
\resizebox{\hsize}{!}{\includegraphics{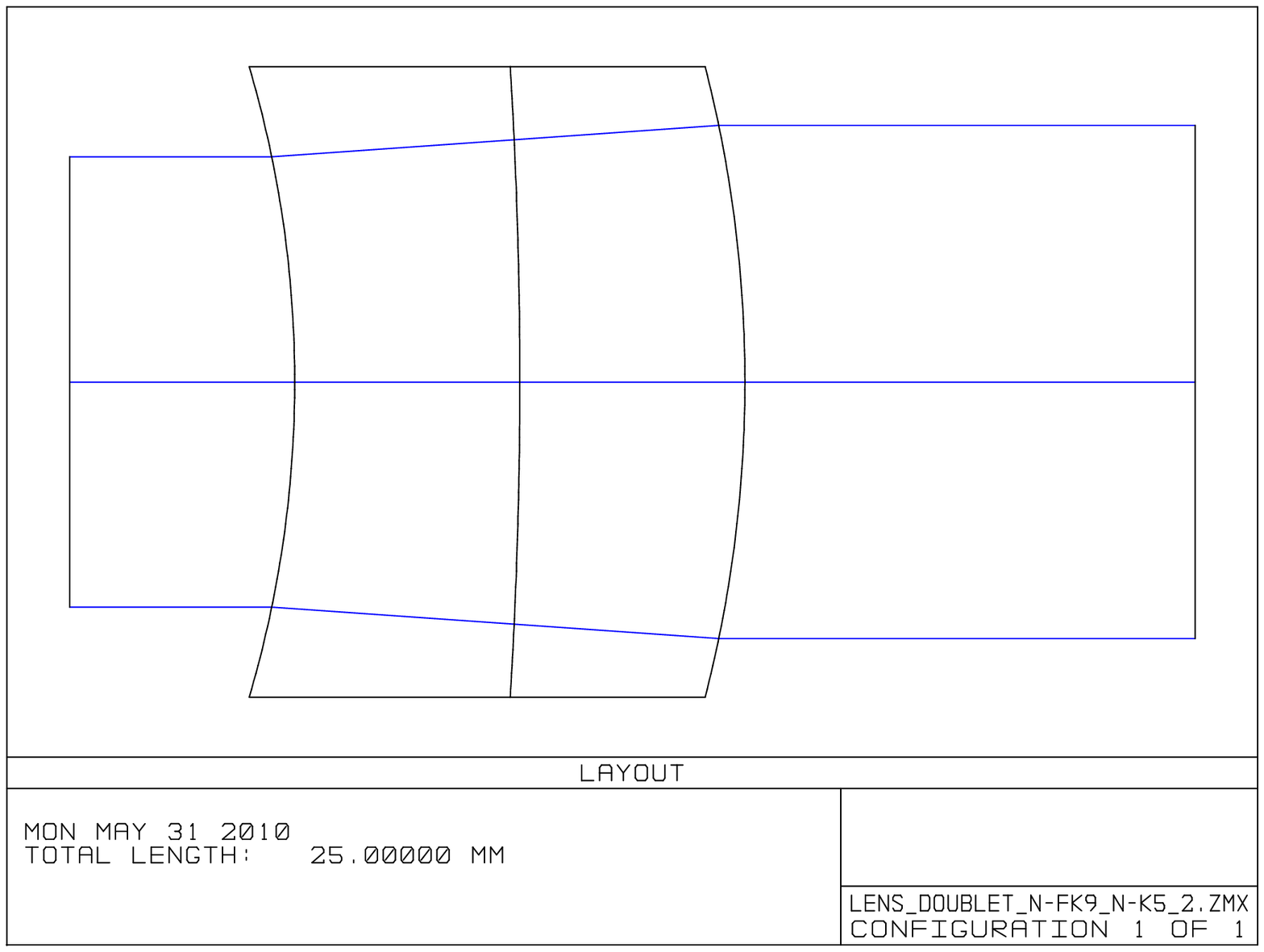}}\\
\resizebox{\hsize}{!}{\includegraphics{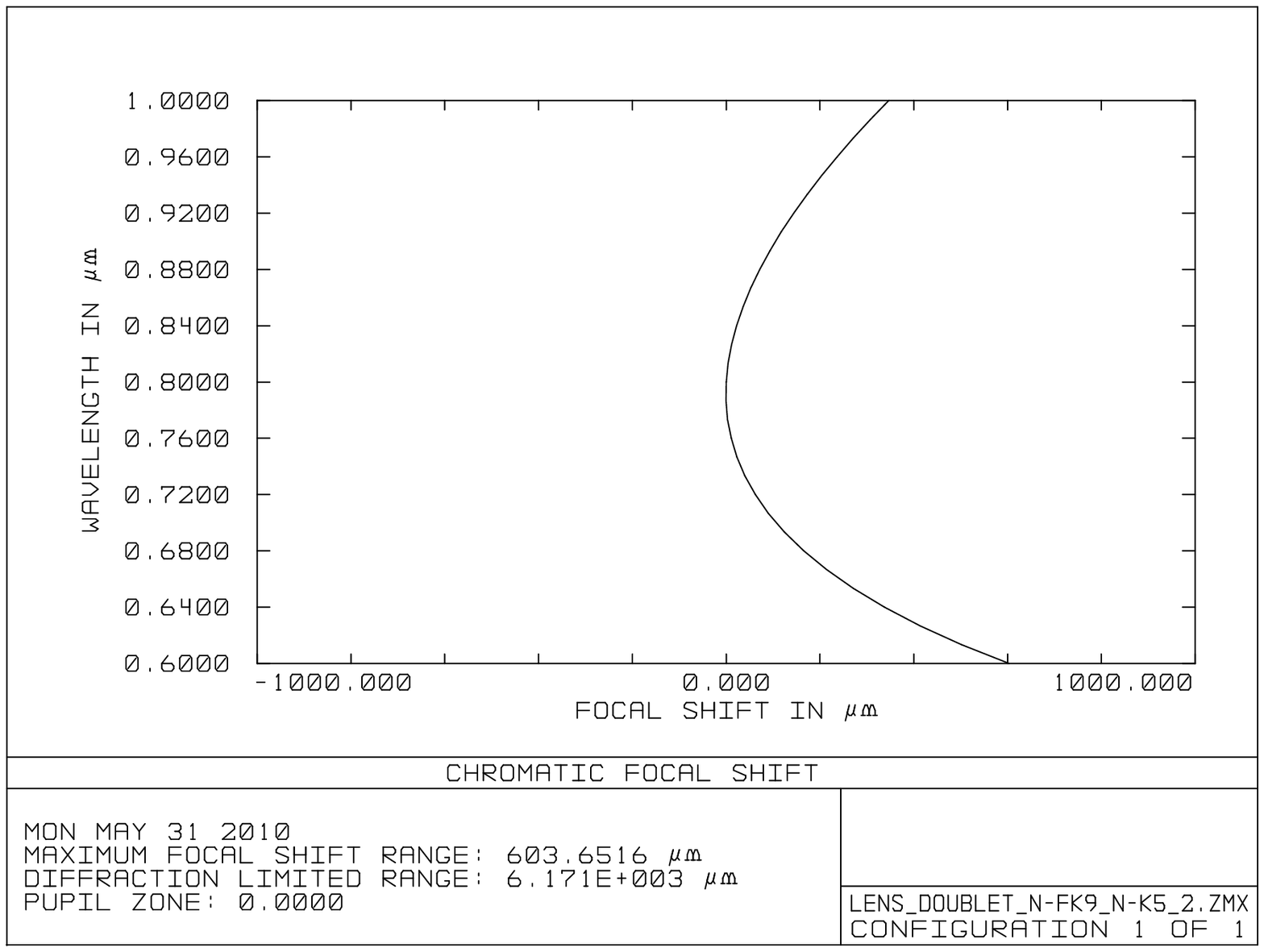}}
\caption{Top: design of a lens doublet combining Schott glass materials N-FK9 and K5. Bottom: focal shift versus wavelength. This result has been obtained assuming a longitudinal focal shift reached in the visible range with our values of $\beta_\lambda$. A 2$^{nd}$ order polynomial chromatic focal shift with a 5\,mm amplitude range has been achieved.} 
\label{fig:lens_doublet}
\end{figure}

The colored apodizer also includes phase apodization $\Phi_w$ with a chromatic parameter $\beta_\lambda$. While grey phase apodization can simply be realized by defocussing the mask, we address here the possibility to produce a chromatic defocus. Our numerical optimization shows that parameter $\beta_\lambda$ varies as $\lambda^2$ for the theoretical model and our physical design, see Figure \ref{fig:thickness_beta_design_parameters}. A possible way to realize a 2$^{nd}$ degree polynomial chromatic defocus consists in using a powerless doublet at the entrance pupil plane. This doublet will present the following characteristics: infinite focal length at $\lambda_0$, a null 1$^{st}$ order chromatic dispersion and a constant and non null 2$^{nd}$ order chromatic dispersion, also called anomalous dispersion. This can be achieved by combining glasses which have the same refractive indices $n_1$ and $n_2$, the same Abbe numbers $\nu_1$ and $\nu_2$, but different partial dispersions. This work follows \citet{2006ApOpt..45.5164D} who has recently realized studies of apochromatic lens design by considering the partial dispersion properties of materials from standard glass catalogs rather than special anomalous-dispersion glasses. We have performed preliminary tests of this approach in the visible range using ZEMAX\textregistered\, ray tracing code and tabulated glass data. Encouraging results for the realization of our phase apodization were found with a combination of Schott glass materials N-FK9 and K5, see Figure \ref{fig:lens_doublet}. Indeed, chromatic behavior in $\lambda^2$ is obtained with this doublet, corresponding to the chromatic defocus expected for the phase apodization of our CAPM coronagraph. Combination of other glass materials are currently under study to achieve adequate chromatic defocus in other spectral bands.

\section{Sensitivity analysis}\label{sec:sensitivity}  
After optimizing the CAPM coronagraph for achromatization, we investigate here several effects that may alter the performance provided by this coronagraphic system. In the following sensitivity analysis, we consider a CAPM coronagraph optimized for a 25\% bandwidth (H-band) observing a point-like source, except for the star angular size study, in the case of a clear circular aperture.

\subsection{Performance criteria}\label{subsec:IDA}
\begin{figure}[!ht]
\centering
\resizebox{\hsize}{!}{\includegraphics{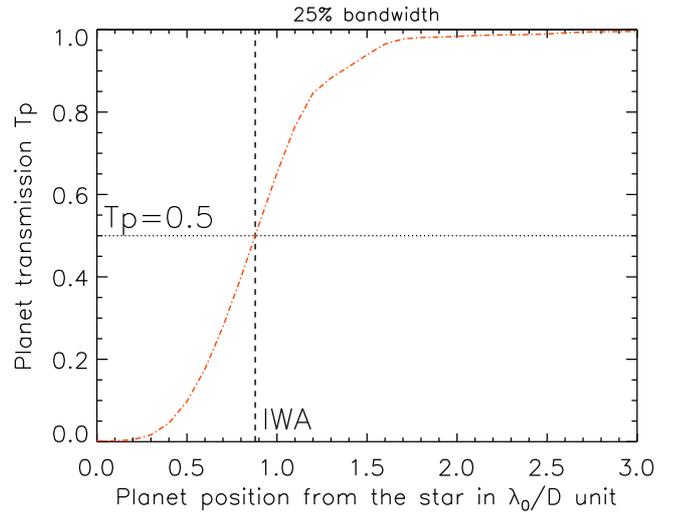}}
\caption{Planet transmission profile achieved with the CAPM coronagraph as a function of the planet position.} 
\label{fig:planet_transmission_profile}
\end{figure}

Contrast performance is judged by the averaged intensity estimated at $5\,\lambda_0/D$ from the main optical axis in the image plane and for an annulus of width $\lambda_0/D$.\\ 
To quantify the capacity to work close to the star, the Inner Working Angle (IWA) is often used, defining the minimum angular distance at which the throughput of a companion is larger than 50\%. Phase mask coronagraphs offer very small IWA, typically $\sim \lambda_0/D$, as recalled for instance by \citet{2008A&A...492..289M}. This is the case for the CAPM coronagraph as illustrated in Figure \ref{fig:planet_transmission_profile}. Unfortunately, IWA is not related to the notion of contrast and therefore, the intensity of close-in companions detectable at a given angular distance from the star cannot be determined with this benchmark.\\
We propose a new criterion, named Inner Detectability Angle (IDA), which represents the compromise between IWA and achievable contrast. IDA is defined as the minimum angular distance $\alpha$ from the main optical axis for which the ratio between the normalized intensity of our coronagraphic star image and the planet transmission flux $T_p$ is less than a given value, here set to $10^{-6}$:
\begin{equation}
\frac{I_D(\alpha)}{I_0} \times \frac{1}{T_p(\alpha)} \leq 10^{-6}\,,
\end{equation}
where $I_0$ denotes the intensity peak value in the absence of coronagraph mask. In Figure \ref{fig:planet_transmission_profile}, we plot the planet transmission profile as a function of the planet position in the presence of a CAPM coronagraph optimized for a 25\% bandwidth. From this curve, we deduce that the planet transmission can reasonably be taken equal to 1 for distances larger than $1.5\,\lambda_0/D$. IDA is expected to be larger than IWA, but estimating IDA allows us to know at which angular distance companions, $10^{-6}$ fainter than their host star, could be detected with our CAPM coronagraph. We study the evolution of IDA for the different cases described below.


\subsection{Effects of aberrations}
Wavefront errors constitute a major issue for stellar coronagraphs as they limit the performance provided by these diffraction suppression systems. This crucial point of aberrations then has to be addressed in the framework of a coronagraphic planet imager. In the case of a ground-based telescope with an XAO system, as in the case of a well-corrected space instrument, the residual quasi-static instrumental aberrations are known to affect the intensity level reached by a downstream coronagraphic device. A performance estimate of the CAPM coronagraph in the presence of aberrations is therefore required to know the sensitivity of our device.  In the following, we describe the behavior of our coronagraphic concept in the presence of low-order aberrations (tip-tilt, defocus, astigmatism, coma, and spherical aberration) since they represent the main contribution in terms of wavefront errors. In addition, we analyze the impact of random aberrations caused by real optical elements on the CAPM coronagraph performance. This aims to determine the limits of our system in the presence of light scattered by the surface roughness of optical elements.\\ 
The low-order aberrations are decomposed here following the Zernike polynomials basis. Their amount is expressed in Zernike rms coefficients and given in $\lambda_0$ unit. The random aberrations considered here are related to the wavefront errors introduced by real optical elements. Good quality optical surfaces are found to exhibit a power spectral density (PSD) function proportional to $\nu^{-2}$ where $\nu$ denotes the spatial frequency in the pupil, as recalled for instance by \citet{1997PASP..109..815R}. To realize our study with random aberrations, we introduce numerically a randomly aberrated phase screen following this PSD, in the entrance pupil plane A of our coronagraph. 

\subsubsection{Qualitative analysis}
\begin{figure*}[!ht]
\centering
\includegraphics[scale=0.325]{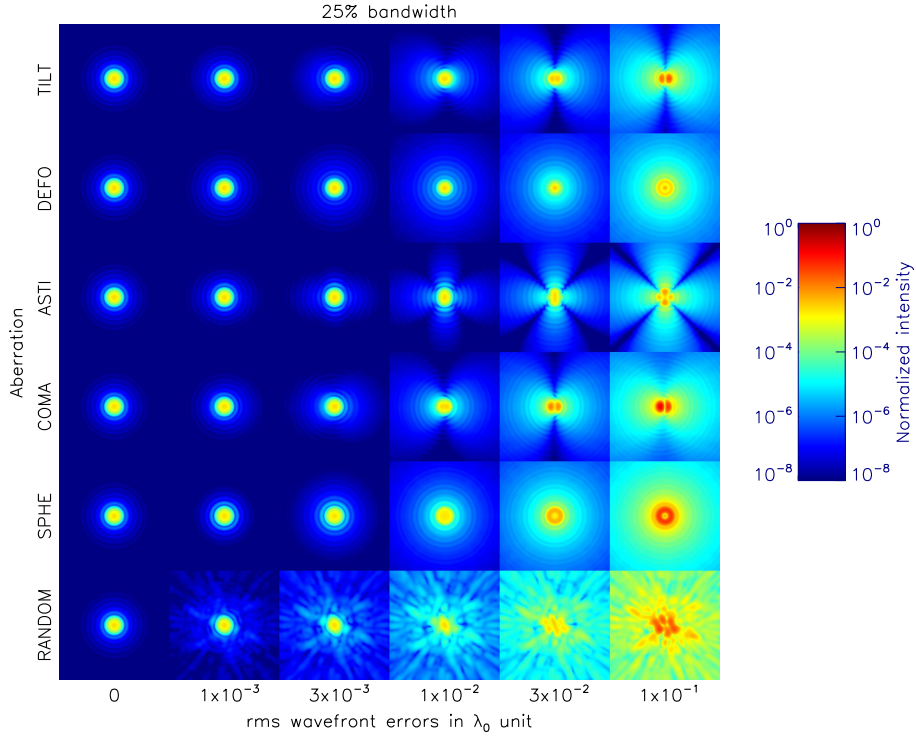}
\caption{Set of coronagraphic images achieved with the CAPM coronagraph and computed numerically for different aberrations. All the images are normalized to the peak of the non coronagraphic image. The pixel scale is the same for all the images and the field of view is represented with a linear range of 22\,$\lambda_0$/D.} 
\label{fig:low-order_aberrations_POLY_PSF_images}
\end{figure*}

Figure \ref{fig:low-order_aberrations_POLY_PSF_images} gives a set of the images theoretically  obtained with the CAPM coronagraph. These frames are represented for different values of rms wavefront errors and each aberration considered here. The shape of the coronagraphic image modifies and evolves quickly as the amount of the rms wavefront errors increases.

\noindent In Figure \ref{fig:aberrations_POLY_profiles}, we draw the radial intensity profiles of the coronagraphic images obtained with the CAPM coronagraph for different rms wavefront error values of defocus (top plot), spherical aberration (middle plot) and random aberration (bottom plot). Through the evolution of the profile with the amount of a given aberration, one can appreciate the variation of the IDA and the intensity level reached with the CAPM coronagraph. A dashed line representing the $10^{-6}$ intensity level set for IDA is also displayed on the plot.\\
We note that while both low-order and random aberrations affect all angular separations, the low-order profiles remain parallel with the non-aberrated profile, following a $\alpha^{-3}$ law, as in the case of the classical Airy pattern. In the random aberrations case, on the other hand, the profile changes towards a $\alpha^{-2}$ profile, corresponding to the PSD profile defined above. We conclude that mechanism leading to performance degradation is different in the two cases. While edge diffraction dominates in the low-order case, wavefront diffraction, as modeled by a sum of phase gratings \citep{1996ifo..book.....G} dominates in the random case.

\subsubsection{Intensity \& IDA}
\begin{figure}[!ht]
\centering
\resizebox{\hsize}{!}{\includegraphics{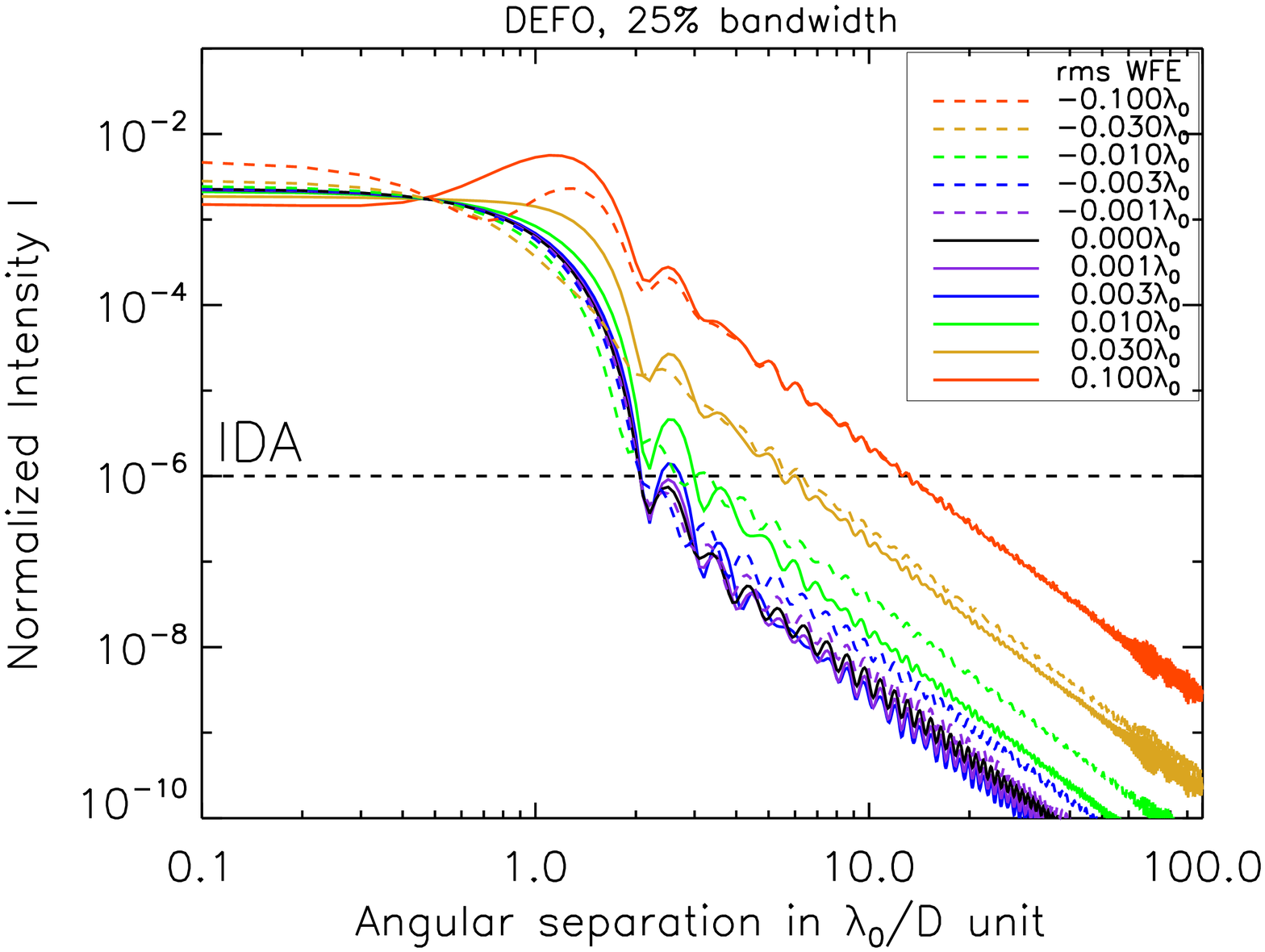}}
\resizebox{\hsize}{!}{\includegraphics{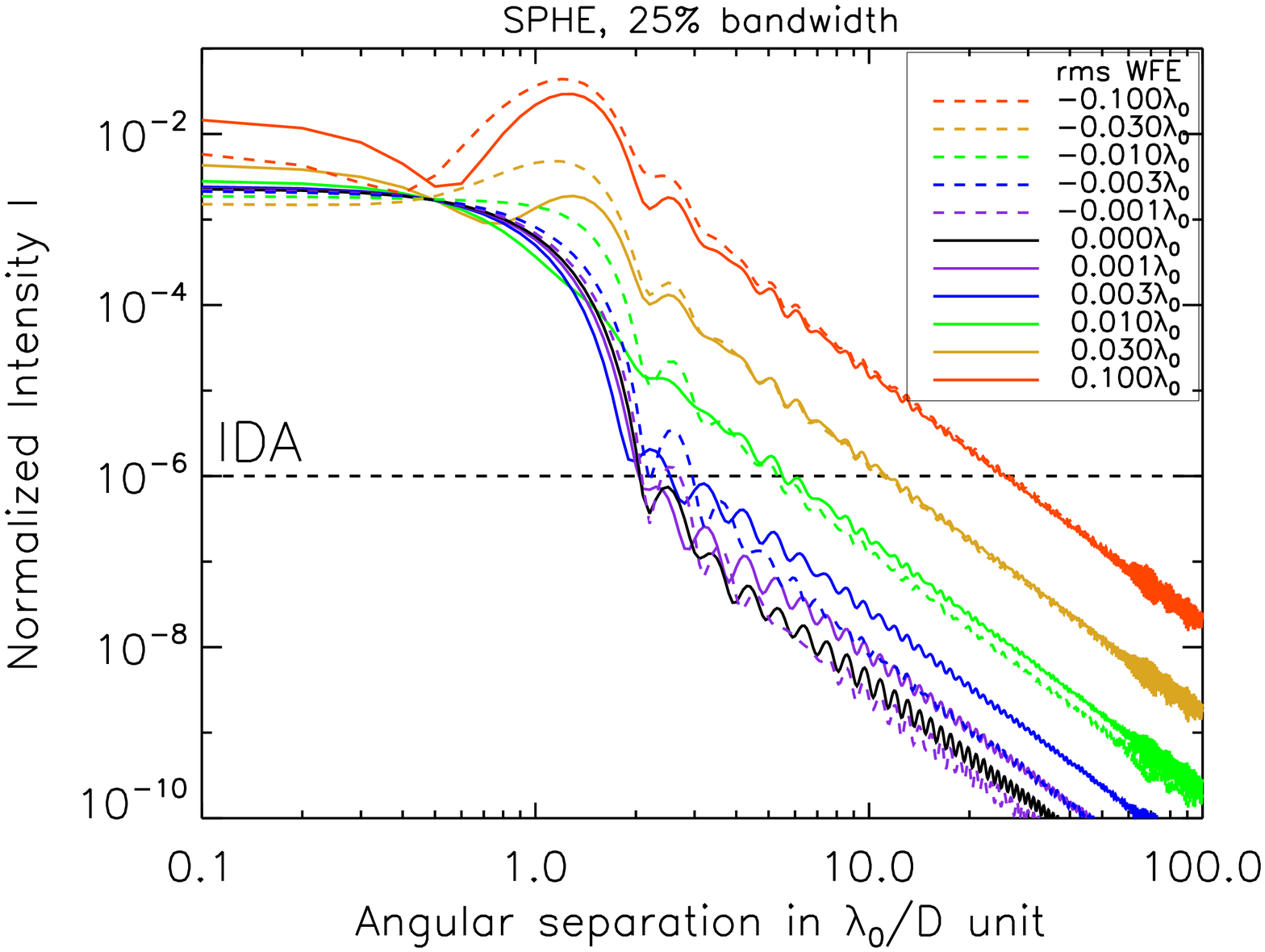}}
\resizebox{\hsize}{!}{\includegraphics{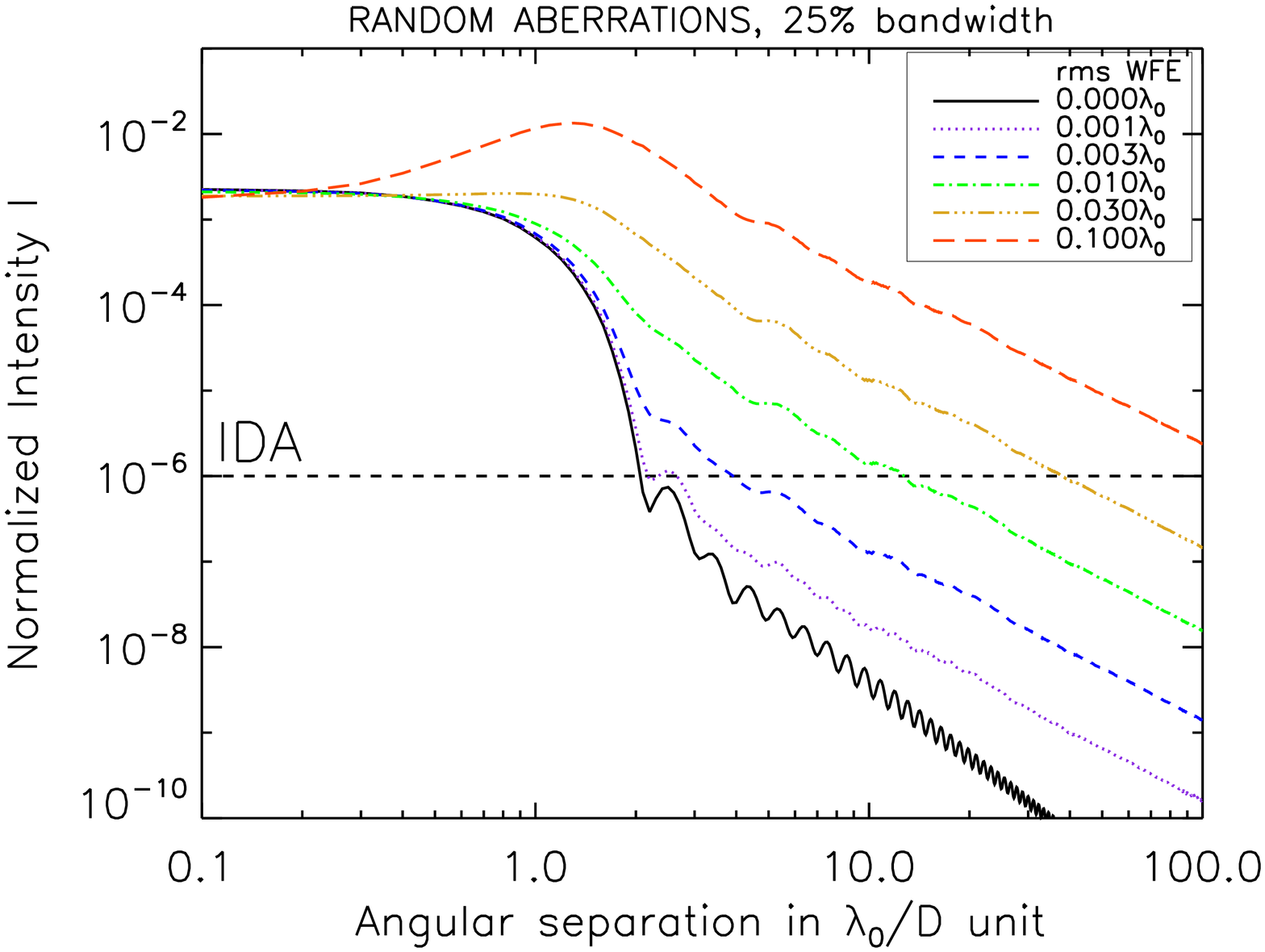}}
\caption{Radial intensity profiles of the coronagraphic images theoretically achieved with the CAPM coronagraph for different rms wavefront errors values of defocus (top plot), spherical aberration (middle) and random aberration with PSD defined above (bottom).} 
\label{fig:aberrations_POLY_profiles}
\end{figure}

\begin{figure}[!ht]
\centering
\resizebox{\hsize}{!}{\includegraphics{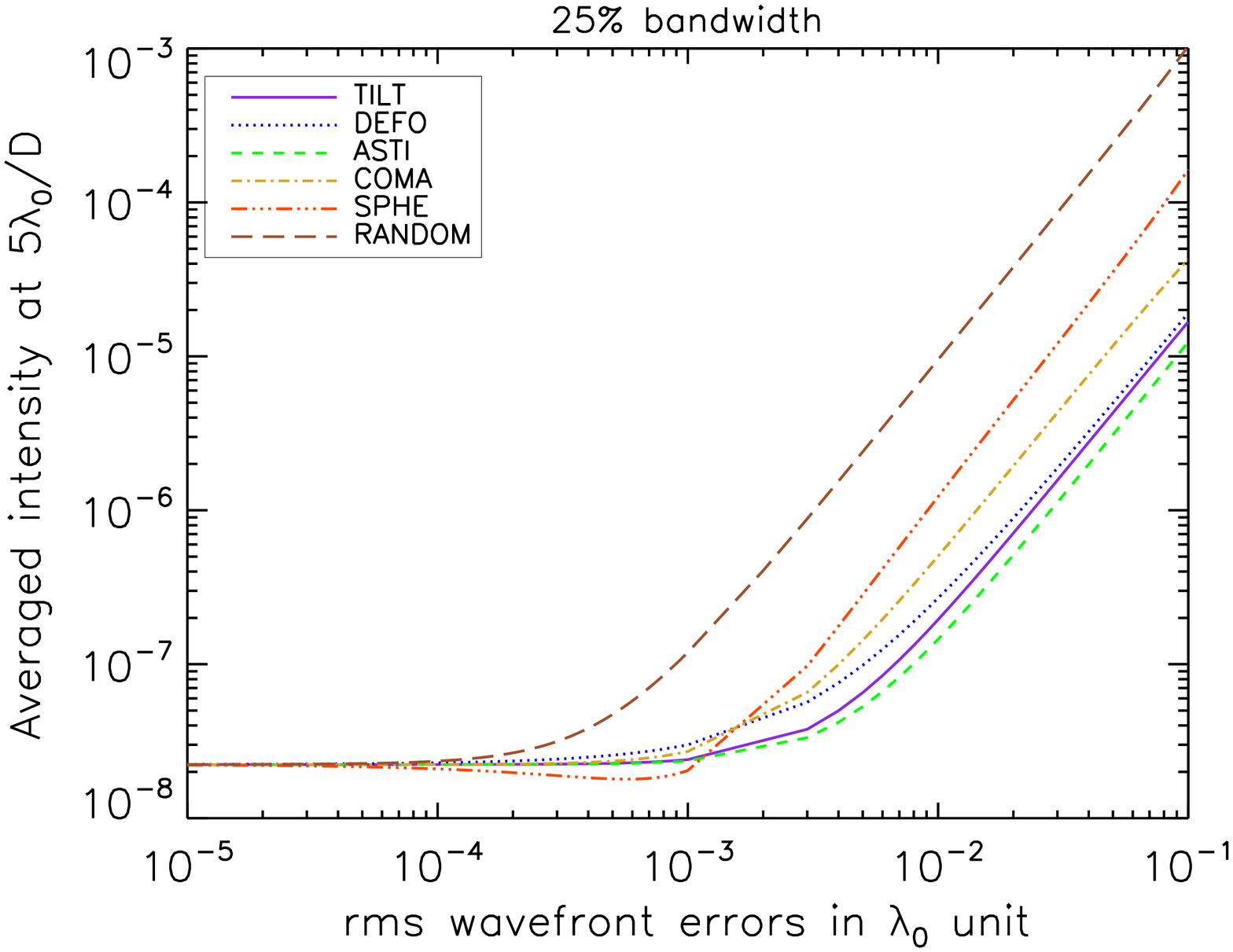}}
\resizebox{\hsize}{!}{\includegraphics{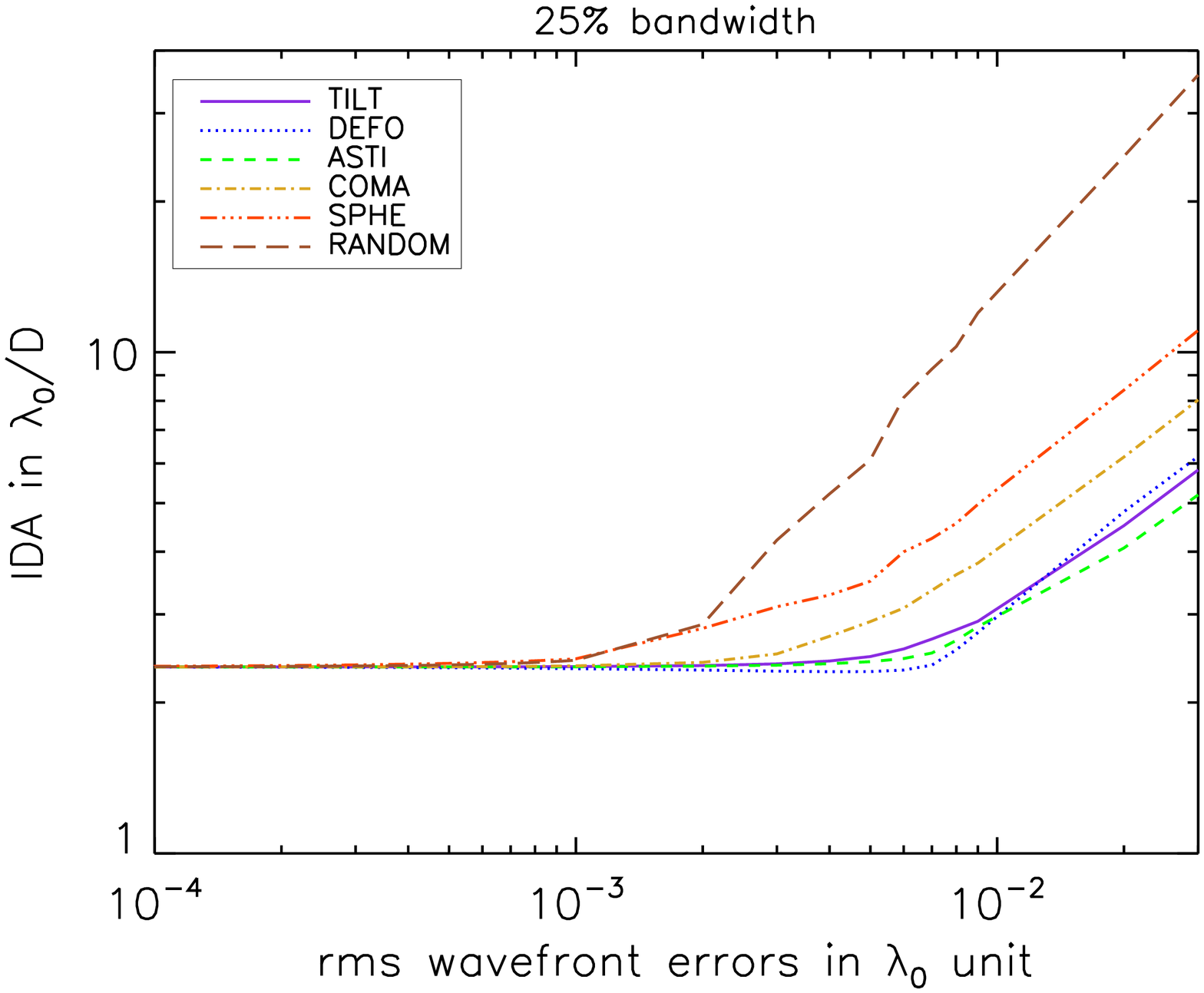}}
\caption{Theoretical averaged intensity at 5\,$\lambda_0/D$ (top plot) and IDA (bottom plot) achieved with the CAPM coronagraph for different aberrations.} 
\label{fig:all_aberrations_POLY}
\end{figure}

In Figure \ref{fig:all_aberrations_POLY} top plot, we represent the averaged intensity at 5\,$\lambda_0/D$ achieved by the CAPM coronagraph as a function of the rms wavefront errors value. The averaged intensity goes beyond a $10^{-7}$ value for rms wavefront errors value larger than $\sim 3\cdot 10^{-3}\,\lambda_0$ and $\sim 9\cdot 10^{-4}\,\lambda_0$ for low-order and random aberrations respectively. At greatest aberration values, the reachable averaged intensity increases approximately as a 2$^{nd}$ degree polynomial function for all the aberrations.\\
Let us note that in the case of a spherical aberration, a small and local improvement of the coronagraph performance is observed for a wavefront error of $\sim 7\cdot 10^{-4}\,\lambda_0$ rms with respect to the reference averaged intensity value in the absence of aberration. We deduce that an additional phase apodization in the form of spherical aberration can slightly improve the contrast gain provided by this device but the investigation of this effect is not within the scope of this paper. However, the gain at $5\,\lambda_0/D$ implies a non desirable increase in intensity of the first Airy ring, see the $10^{-3}\,\lambda_0$ rms profile in Figure \ref{fig:aberrations_POLY_profiles} middle plot.\\
To complete the analysis, we represent the IDA theoretically achieved with the CAPM coronagraph as a function of the rms wavefront errors value, see Figure \ref{fig:all_aberrations_POLY} bottom plot. IDA remains below 3\,$\lambda_0/D$ for rms wavefront errors lower than $3 \cdot 10^{-3}\,\lambda_0$ and $2 \cdot 10^{-3}\,\lambda_0$ for the low-order and random aberrations respectively. At larger values, the IDA evolves approximately as a 2/3 and a second degree polynomial function for the low-order and random aberrations respectively.

\subsubsection{Discussion}
The intensity level and IDA provided by CAPM coronagraph are all the more altered as the aberration order is higher. These results show that at a given value of rms aberration, the sensitivity of our coronagraph increases with the order of the aberration. They also underline the necessity of working with very high quality optical elements, typically better than $10^{-3}\,\lambda_0$ rms wavefront error, to enable high performance of the CAPM coronagraph for a very high contrast imaging as well as a very small IDA.\\
Several studies of sensitivity to low-order aberrations have been performed for different kinds of coronagraph \citep[e.g][]{2005ApJ...628..474S,2005ApJ...634.1416S,2006ApJ...652..833B,2010SPIE.7739E..33M}. No general behavior of the coronagraphs to low-order aberrations emerge from the previous analysis since each concept presents a specific response to these wavefront errors. In addition and to the best of our knowledge, no advanced study has been reported on the response of circular phase mask coronagraphs to low-order aberrations.


\subsection{Effects of mask imperfections}
\subsubsection{Mask roughness}
Effects of the mask roughness have to be analyzed since they could alter the performance of the CAPM coronagraph. Many glass substrates with excellent flatness (better than $\lambda/20$, see for instance Newport \textregistered 10QW40-30 glass) are available and useful for our DZPM manufacturing. To lead our study, we therefore assume a perfectly flat substrate of the DZPM except where the phase disk and annulus are machined into it. So, only effects of surface roughness present inside the mask area of diameter $d_2$ have been considered.\\
In Figure \ref{fig:mask_imperfections_POLY_profiles} top plot, we draw the radial profile of the coronagraphic image for different values of rms mask roughness. All the coronagraphic profiles follow a $\alpha^{-3}$ law as that of the Airy diffraction pattern does and their intensity level increases with the amount of rms mask roughness.\\
Following this result, we also represent the averaged intensity at $5\,\lambda_0/D$ reached by the CAPM coronagraph as a function of the rms mask roughness, see Figure \ref{fig:mask_alleffects} top plot. It can be noticed that a $10^{-7}$ intensity level is provided with the CAPM coronagraph for rms mask roughness values lower than $\sim 0.12\,\lambda_0$. Beyond this value, the intensity increases and follows the behavior of a 2$^{nd}$ degree polynomial function until it reaches a plateau at 2\,$\cdot 10^{-5}$, corresponding to the non-coronagraphic case, for roughness greater than 1.00\,$\lambda_0$ rms.\\
In addition, the IDA profile is displayed in Figure \ref{fig:mask_alleffects} bottom plot. A $3\,\lambda_0/D$ IDA is preserved for rms mask roughness value lower than 0.15$\,\lambda_0$. For larger roughness values, the IDA increases and follows a 1$^{st}$ degree polynomial function, again reaching a plateau at 1.00\,$\lambda_0$ rms.\\
For 0.12\,$\lambda_0$ rms mask roughness, the corresponding values are 198.0\,nm and 76.0\,nm at $\lambda_0=1650$\,nm (H-band) and  at $\lambda_0=633$\,nm (visible) respectively. As a reference, the roughness obtained with our second-generation RRPM is 0.8\,nm rms \citep{2010A&A...509A...8N}. Since the same accuracy can be obtained for the manufacturing of DZPM as for that of the RRPM, we can deduce that mask roughness will not constitute a major issue for the CAPM coronagraph.

\subsubsection{Mask transition zone width}
\begin{figure}[!ht]
\centering
\resizebox{\hsize}{!}{\includegraphics{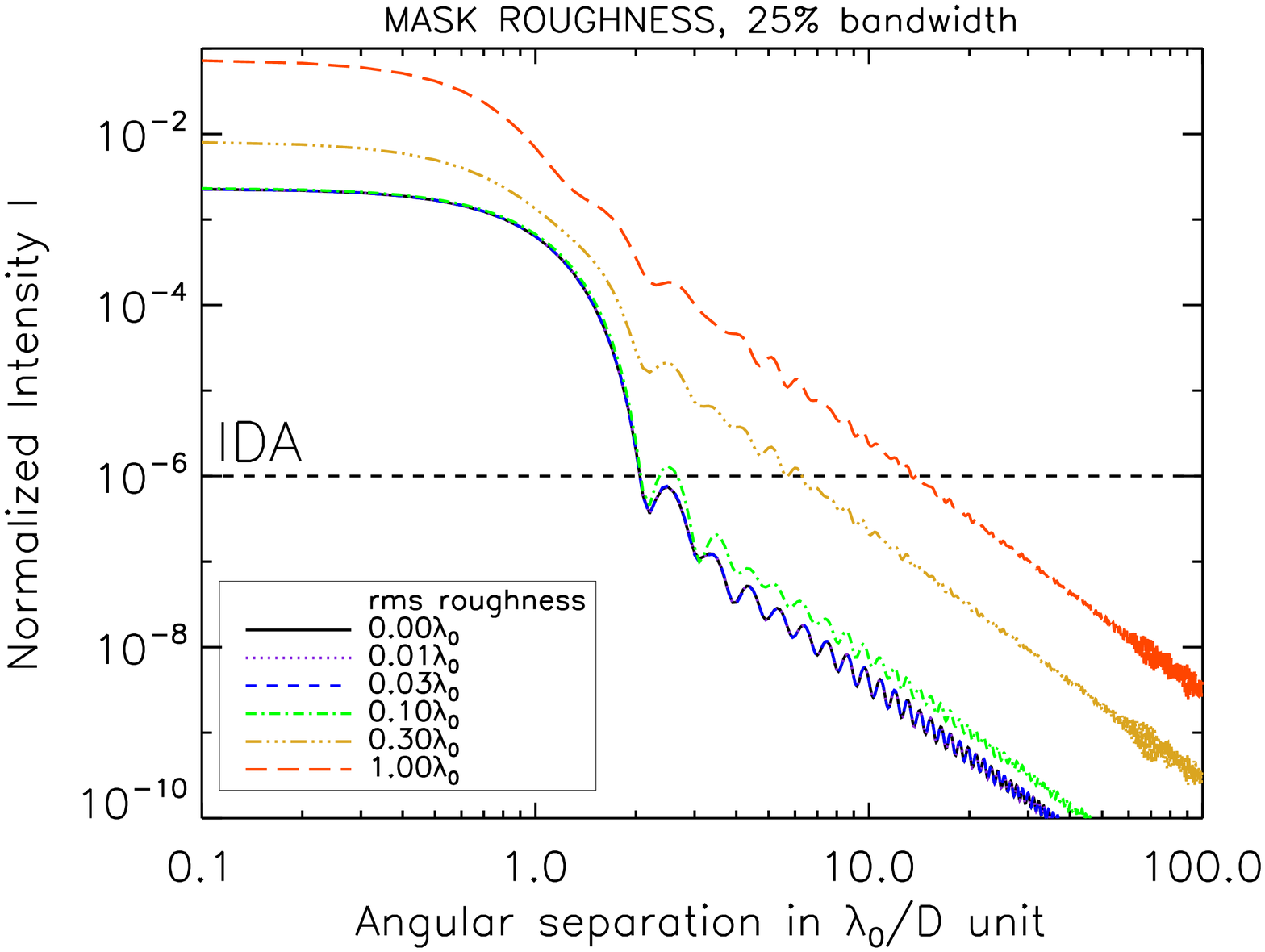}}
\resizebox{\hsize}{!}{\includegraphics{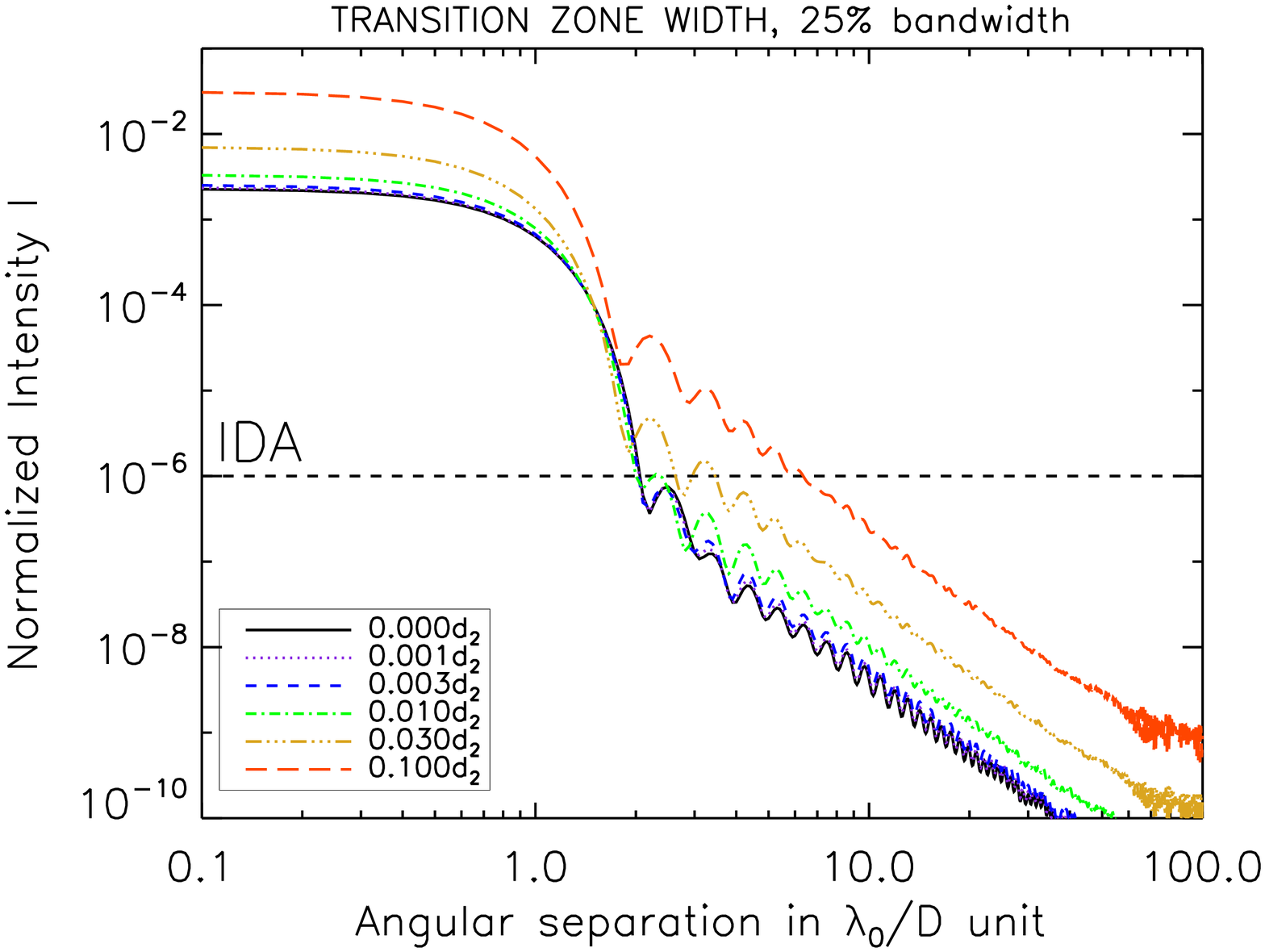}}
\caption{Radial intensity profiles of the coronagraphic images theoretically achieved with the CAPM coronagraph for different rms values of mask roughness (top plot) and for different sizes of mask transition width, expressed in DZPM diameter $d_2$ (bottom).} 
\label{fig:mask_imperfections_POLY_profiles}
\end{figure}

\begin{figure}[!ht]
\centering
\resizebox{\hsize}{!}{\includegraphics{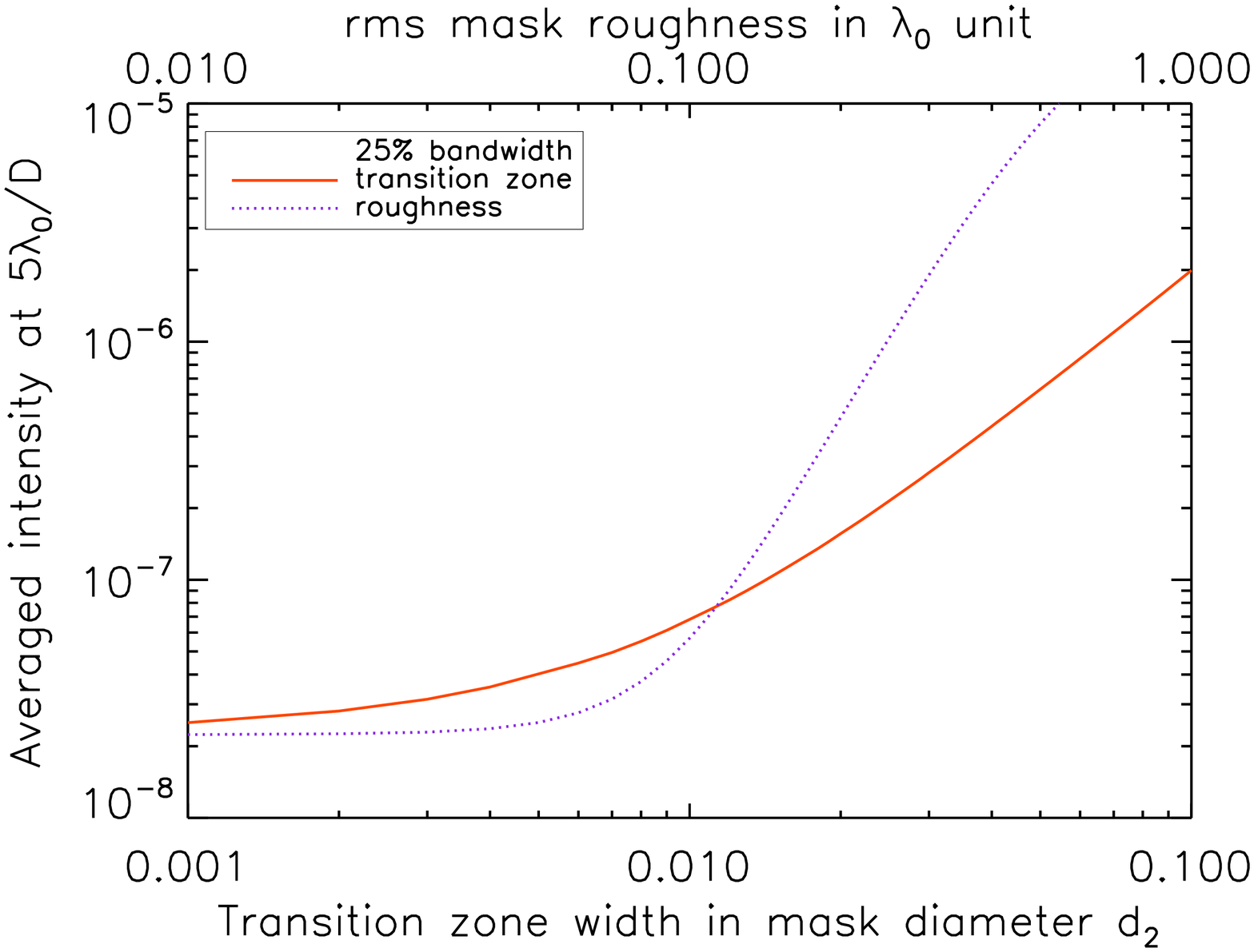}}
\resizebox{\hsize}{!}{\includegraphics{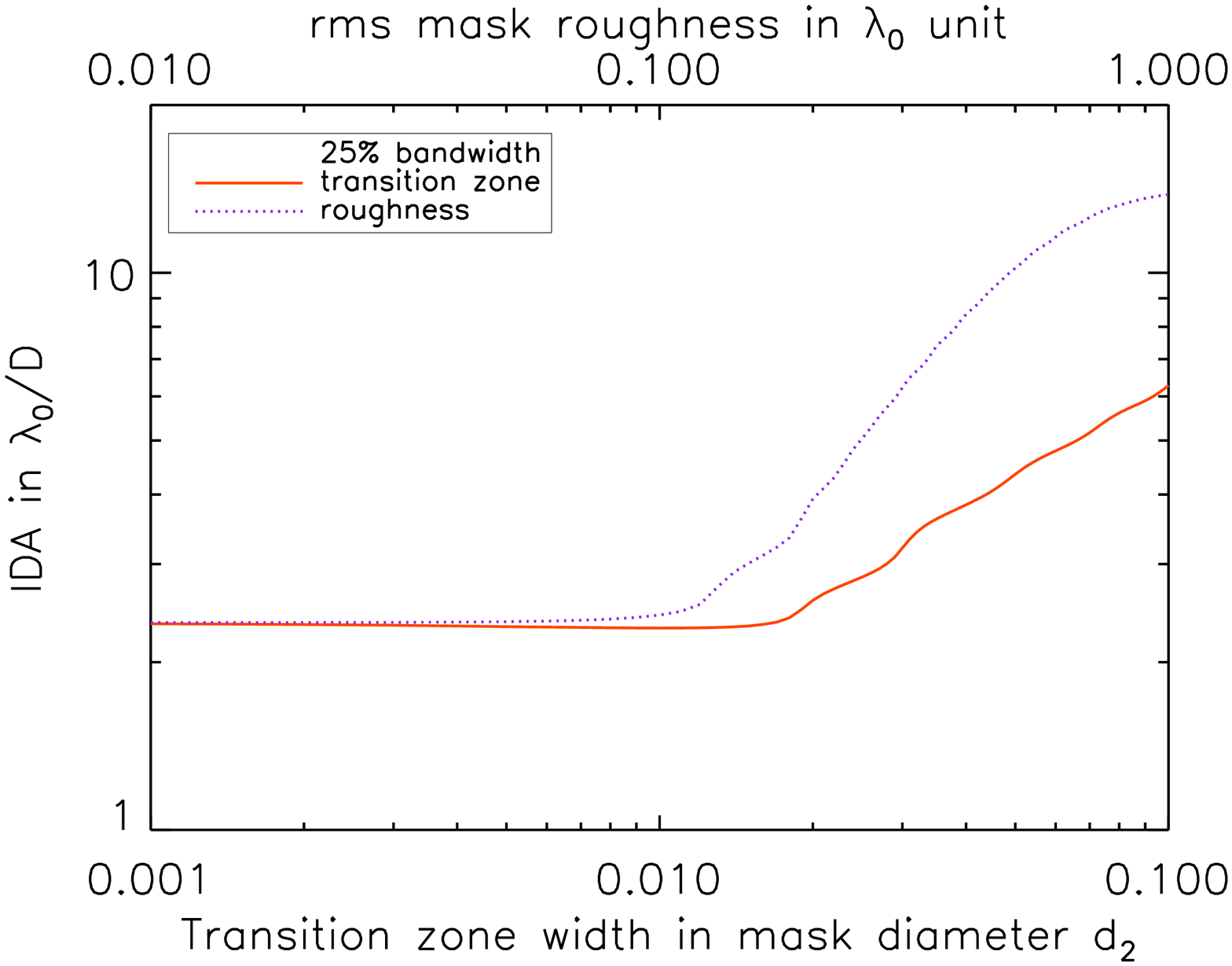}}
\caption{Theoretical averaged intensity (top plot) and IDA (bottom plot) achieved with the CAPM coronagraph as a function of the size of the mask transition zone (bottom axis) and the mask roughness (top axis).} 
\label{fig:mask_alleffects}
\end{figure}

The DZPM is theoretically a hard-edge mask. However in practice, some transition zone widths are found in a manufactured prototype as noticed with the second-generation RRPM \citep{2010A&A...509A...8N}. We propose to study the impact of the DZPM transition width on the CAPM coronagraph performance by replacing vertical transitions by slopes in the mask shape. The same width is assumed here for the inner and outer parts of the DZPM and expressed in mask outer part diameter $d_2$. The transition zone is illustrated in Figure \ref{fig:DZPM_drawing}.\\
Radial profiles of the coronagraphic images are displayed for different mask transition widths in Figure \ref{fig:mask_imperfections_POLY_profiles} bottom plot. All the profiles evolve as a $3^{rd}$ degree polynomial function like the Airy diffraction pattern. As expected, the intensity values of the coronagraphic image increase as the DZPM transition zone width becomes larger.\\  
In Figure \ref{fig:mask_alleffects} top plot, we display the averaged intensity at $5\,\lambda_0/D$ provided by the CAPM coronagraph as a function of the DZPM transition zone width. An intensity level of $10^{-7}$ is preserved for a mask transition width smaller than 0.014\,$d_2$. At larger widths, the averaged intensity increases but slower than a 2$^{nd}$ degree polynomial function.\\
Figure \ref{fig:mask_alleffects} bottom plot shows the evolution of IDA achieved with the CAPM coronagraph as a function of the mask transition width. We can notice that IDA remains below $3\,\lambda_0/D$ for a mask transition width lower than $0.028\,d_2$. Beyond this value, IDA evolves as a $1/2$ degree polynomial function, showing the low sensitivity of this criteria to an increase of the mask transition width.\\
As a summary, a 1\% mask transition width allows us to preserve a $10^{-7}$ intensity level and good IDA with the CAPM coronagraph. This value is similar to that measured for the second-generation RRPM which was less than 1\% of the mask diameter \citep{2010A&A...509A...8N} and below 1\,$\mu$m. The same accuracy is expected concerning the manufacturing of the DZPM. A 1\% mask transition width corresponds physically to $1\,\mu$m for a 100$\,\mu$m DZPM diameter. Hence, the focal ratio $F$ of the coronagraphic layout will range as follows: $F \sim 61$ at $\lambda=1.65\,\mu$m, $F \sim 100$ at $\lambda=1\,\mu$m and $F \sim 200$ at $\lambda=0.5\,\mu$m. Below a 1\,$\mu$m wavelength, $F$ constitutes a severe constrain on the design of a planet imager working in the visible and in which a CAPM coronagraph could be inserted. To overcome focal ratio issues in the visible spectral range, manufacturing DZPM with a diameter and mask transition width smaller than 100$\,\mu$m and 1\,$\mu$m respectively is required.

\subsection{Lyot stop size and misalignment}
\subsubsection{Lyot stop size}
\begin{figure}[!ht]
\centering
\resizebox{\hsize}{!}{\includegraphics{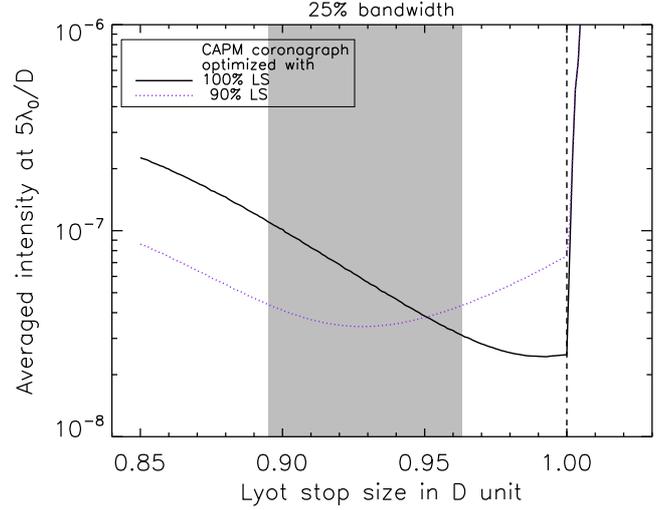}}
\caption{Theoretical averaged intensity at $5\,\lambda_0/D$ achieved with the CAPM coronagraph as a function of the Lyot stop (LS) size. The dashed-line represents the edge of the geometrical relayed pupil. The grey area represents the tolerance range, defined in the text, for the CAPM coronagraph optimized with 90\% LS.} 
\label{fig:lyotstop}
\end{figure}

Pupil undersizing in the Lyot plane is often required in the context of instrument design to allow alignment and manufacturing tolerance. The sensitivity of the CAPM coronagraph to this parameter has therefore to be taken into account. We analyze the impact of a Lyot Stop (LS) size reduction on the contrast provided by the CAPM coronagraph. Two CAPM coronagraphs optimized with a 90\% and a 100\% LS are considered here.\\
In Figure \ref{fig:lyotstop}, the averaged intensity at $5\,\lambda_0/D$ achieved by the CAPM coronagraph is represented as a function of the pupil size for both designs. In each case, the best contrast provided by the CAPM coronagraph is almost obtained for the LS size for which the coronagraphic device has been optimized. However, the behavior of the CAPM coronagraphs differs from each other. In the case of the CAPM coronagraph optimized with a 100\% LS, the intensity level increases from 2\,$\cdot 10^{-8}$ to 2\,$\cdot 10^{-7}$ as we reduce the size of the LS from 1\,$D$ to 0.85\,$D$. On the opposite, in the case of the CAPM coronagraph optimized with a 90\% LS, the intensity level varies slowly from 3\,$\cdot 10^{-8}$ to 5\,$\cdot 10^{-8}$ for LS sizes diverging from 0.93\,$D$. A grey area in the plot underlines the range of Lyot stop sizes for which weak fluctuations of the coronagraph performance are observed. The CAPM coronagraph optimized with 90\% LS is less sensitive to Lyot Stop reduction than that optimized with 100\% LS.\\
So, pupil undersizing weakly alters the CAPM coronagraph performance if this coronagraphic system has been optimized considering the Lyot Stop size. Let us notice that in the case of a Lyot stop size larger than the geometric pupil diameter, the performance of the CAPM coronagraph decreases quickly as expected since the rejected starlight is found there. 

\subsubsection{Lyot stop misalignment}
\begin{figure}[!ht]
\centering
\resizebox{\hsize}{!}{\includegraphics{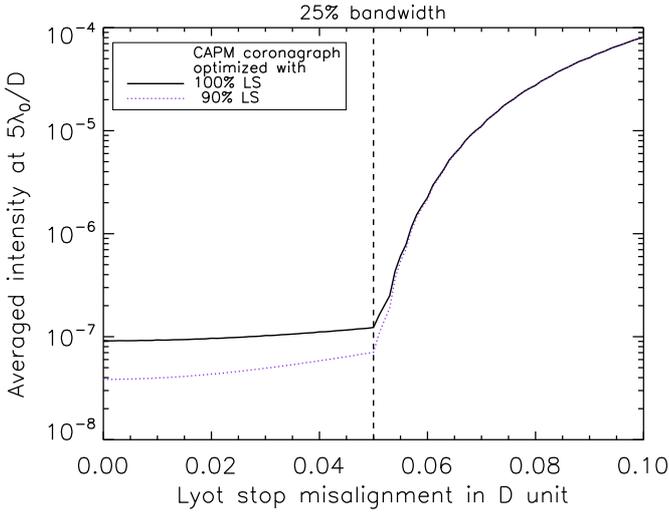}}
\caption{Theoretical averaged intensity at $5\,\lambda_0/D$ achieved with the CAPM coronagraph and a 90\% Lyot stop, as a function of the Lyot stop misalignment. The dashed-line represents the edge of the geometrical relayed pupil. These results are obtained in the presence of a clear circular aperture.} 
\label{fig:pupilshear}
\end{figure}

Lyot stop misalignment has been identified as one of the issues for the coronagraphs on VLT-SPHERE. A misalignment less than 0.2\% of the pupil diameter has been imposed to the selected coronagraphs for the instrument requirements \citep{2006Msngr.125...29B}. In the following, we analyze the impact of the Lyot Stop misalignment with respect to the geometric pupil on the CAPM coronagraph performance. The LS size is taken here equal to 90\% the relayed pupil diameter $D$. Once again, we consider two CAPM coronagraphs: optimized with a 90\% and 100\% LS.\\     
In Figure \ref{fig:pupilshear}, the averaged intensity at $5\,\lambda_0/D$ provided by the DZPM coronagraph is represented as a function of the Lyot Stop misalignment. In the absence of pupil misalignment, the intensity level achieved by the CAPM coronagraph optimized with a 90\% LS and 100\% LS are about 4\,$\cdot 10^{-8}$ and 8\,$\cdot 10^{-8}$ respectively. The result achieved by the CAPM coronagraph optimized to 90\% LS is better than that optimized to 100\% LS since we are facing a 90\% LS in the coronagraphic scheme as mentionned above.\\
As the pupil misalignment increases up to 0.05\,$D$, the intensity level provided by the CAPM coronagraph does not really change significantly. Clearly, the CAPM coronagraph is not sensitive to pupil misalignment for a LS remaining completely inside the geometric pupil.\\
Beyond a 0.05\,$D$ pupil misalignment, the Lyot Stop starts to go outside the geometric pupil image where most of the rejected starlight is found. It results in a very high increase of the averaged intensity and an expected dramatic diminution in performance for the CAPM coronagraph. 

\subsection{Star angular size}
A Sun-like star at 10\,pc is 1 milliarcsec across, corresponding to 0.024\,$\lambda_0/D$ and 0.125\,$\lambda_0/D$ in H-band for a 8m-class telescope and the E-ELT respectively. Since phase mask coronagraphs are quite sensitive to star size \citep{2006ApJS..167...81G}, we analyze the impact of this latter on the performance of our concept.\\
In Figure 25, we plot the azimuthally averaged intensity profiles with the CAPM coronagraph for different star angular radii $\theta$. All the profiles follow a $\alpha^{-3}$ law like the Airy diffraction pattern and the intensity values increase with $\theta$, as expected. In Figure 26 top and bottom plots, we represent the averaged intensity at 5\,$\lambda_0/D$ and IDA respectively as a function of the star angular radius. In the top plot, the intensity level goes beyond $10^{-7}$ for $\theta$ larger than 0.018\,$\lambda_0/D$. At larger angular radii, the intensity level increases as $\theta^2$. In bottom plot, a 3\,$\lambda_0/D$ IDA is preserved for $\theta$ lower than 0.025\,$\lambda_0$/D. At larger values, IDA evolves as $\theta$.\\
As a consequence, the finite size of the sun-like star at 10\,pc has little impact on the CAPM coronagraph when observed through an 8\,m-class telescope. However, for a 40\,m-class telescope, the finite size of this star clearly limits the coronagraphic performance. These results confirm the limitations observed by \citet{2006ApJS..167...81G} about the circular phase mask coronagraphs for the direct imaging of extrasolar terrestrial planets. In order to use the CAPM coronagraph at its optimal performance with an ELT, it is therefore more appropriate to observe more compact or more distant stellar objects.

\begin{figure}[!ht]
\centering
\resizebox{\hsize}{!}{\includegraphics{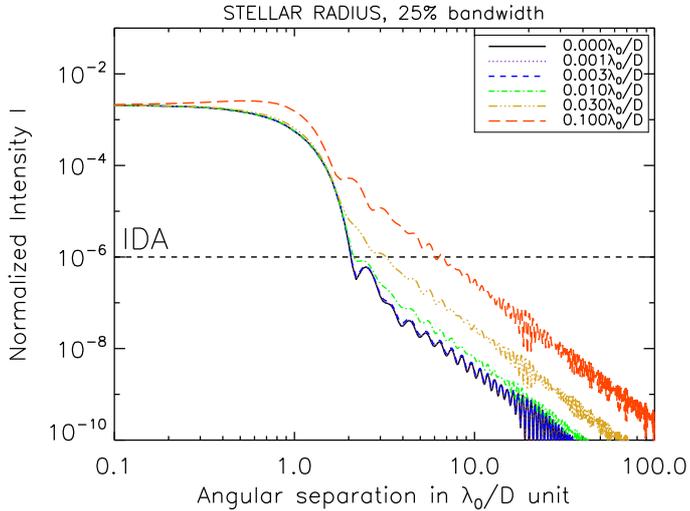}}
\caption{Radial intensity profiles of the coronagraphic images theoretically achieved with the CAPM coronagraph for different stellar radii.} 
\label{fig:stellar_radii_POLY_profiles}
\end{figure}


\begin{figure}[!ht]
\centering
\resizebox{\hsize}{!}{\includegraphics{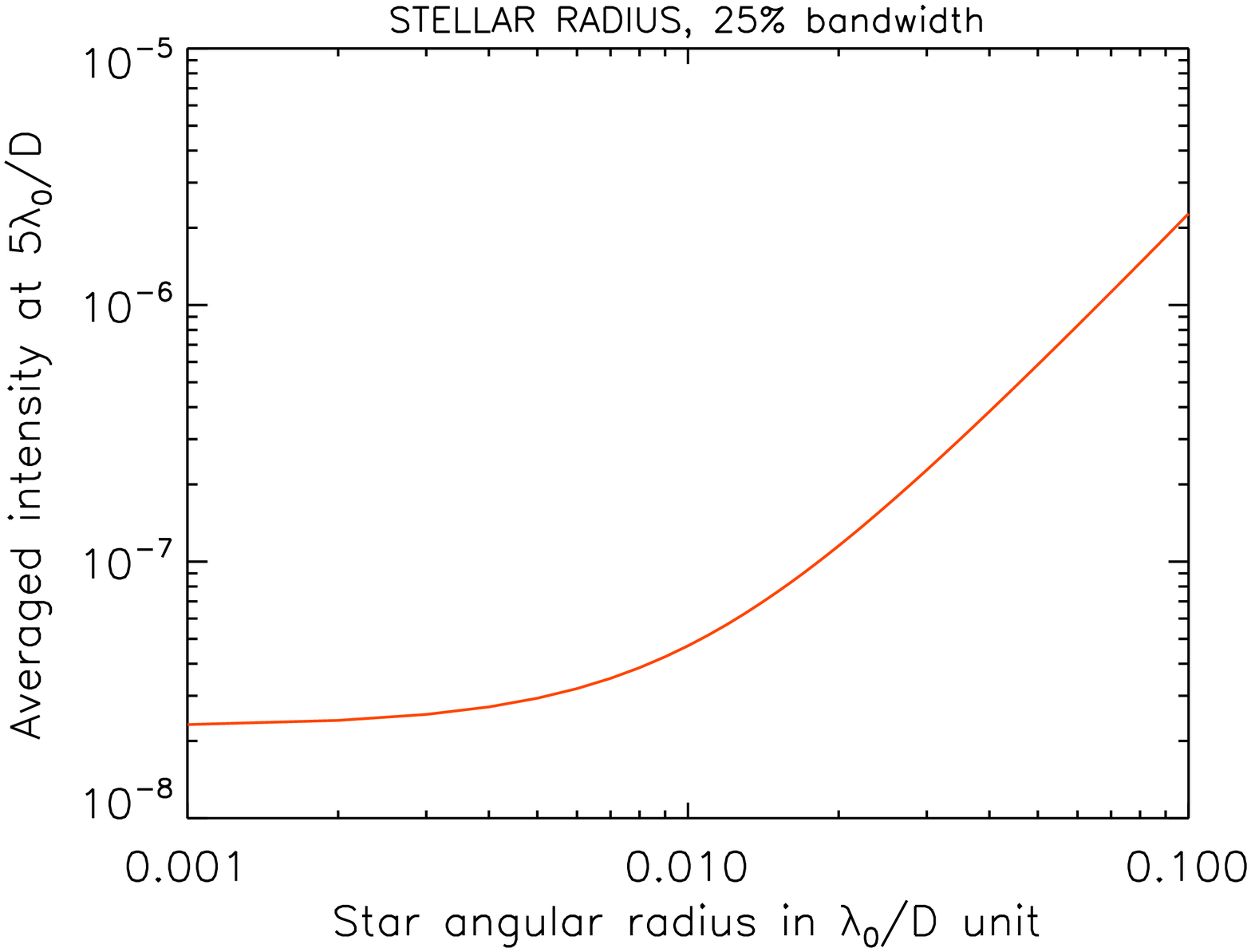}}
\resizebox{\hsize}{!}{\includegraphics{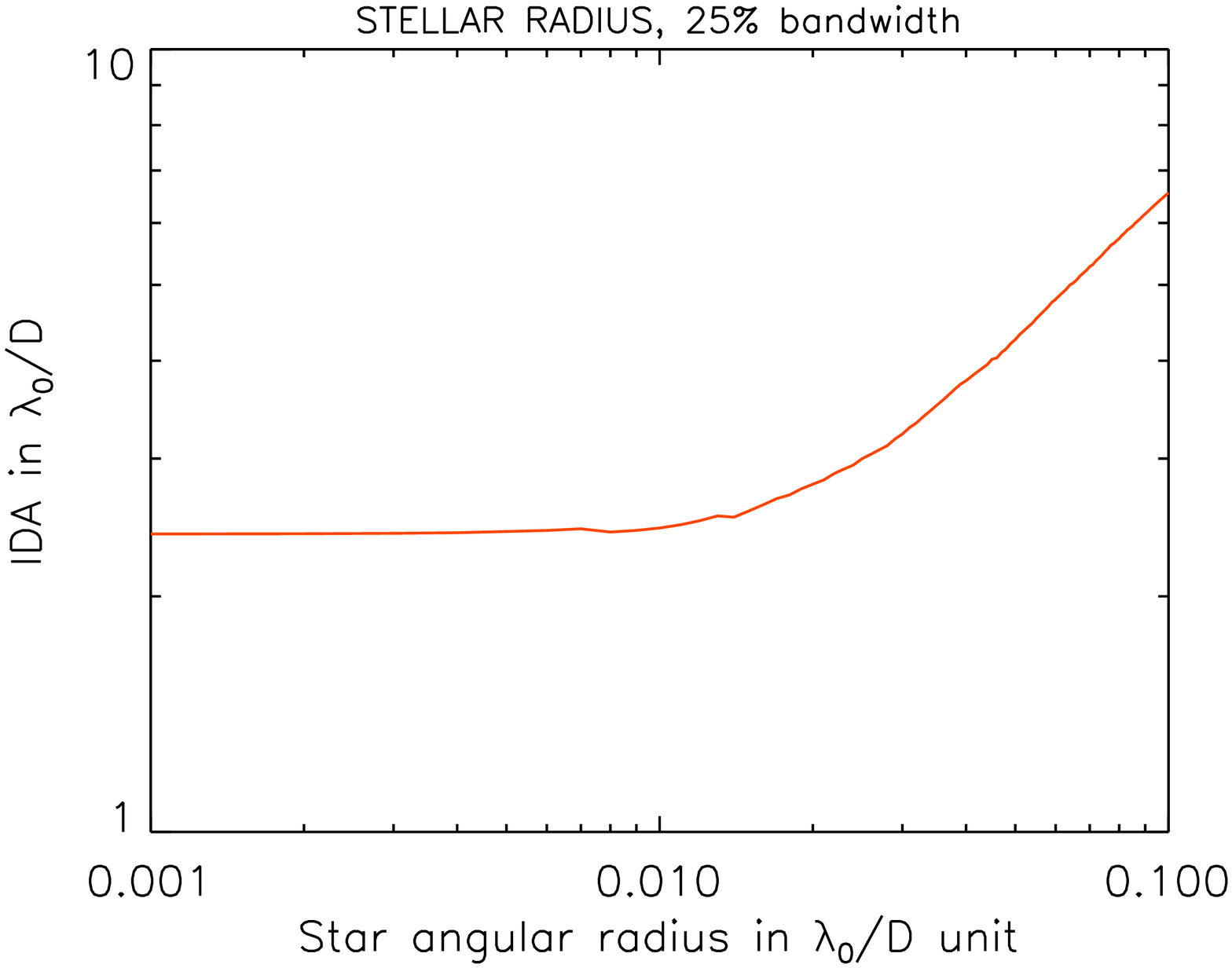}}
\caption{Theoretical averaged intensity at 5\,$\lambda_0/D$ (top plot) and IDA (bottom plot) achieved with the CAPM coronagraph for different stellar radii.} 
\label{fig:stellar_radii_POLY}
\end{figure}

\section{Conclusion}\label{sec:conclusion}  
The CAPM coronagraph is an original concept composed of a colored pupil apodization, a DZPM in the following focal plane and a Lyot Stop in the re-imaged pupil plane. This diffraction suppression system constitutes an improved version of the DZPM coronagraph \citep{2003A&A...403..369S} in which the grey apodization has been replaced by a colored one. This change allows us to increase the achromaticity of circular phase mask coronagraphs; indeed, a 2.5\,$\magn$ contrast gain for the CAPM coronagraph is achieved with respect to the DZPM coronagraph. In addition, intensity levels of 2.2$\cdot 10^{-7}$ and 2.2$\cdot 10^{-8}$ at 3\,$\lambda_0/D$ and 5\,$\lambda_0/D$ respectively from the observed bright star are expected with the CAPM coronagraph, in the case of a clear aperture and a 25\% bandwidth. This performance situates the concept in an excellent position with respect to similar concepts reviewed in the introduction to this paper.\\ 
Manufacturing aspects of the colored apodizer have also been addressed, leading to a promising solution for the realization of our CAPM coronagraph using a nickel thin film for the colored amplitude apodizer. Optimizing the metallic layer thickness and mask parameters, we obtained the ultimate performance of the CAPM coronagraph. We consider to use a lens doublet combining two glasses with the same refractive index, the same Abbe number and different anomalous dispersion to produce colored phase apodization. An encouraging solution has been found for the visible range and studies are on-going to address other spectral bands with the same concept.\\
A sensitivity analysis of the CAPM coronagraph to different errors has also been performed, leading to the determination of the error level which can be accepted before the performance of the device is altered. We find in particular that for a given amount of rms wavefront errors, the concept is less sensitive to low-order than to high-order aberrations. While an intensity level at 5\,$\lambda_0/D$ of $10^{-7}$ is ensured for about $6\cdot 10^{-3}\,\lambda_0$ of tilt or defocus, the same intensity level requires spherical aberration less than $3\cdot 10^{-3}\,\lambda_0$. In the case of random aberrations with a $\nu^{-2}$ PSD distribution, this intensity level requires wavefront errors below $10^{-3}\,\lambda_0$. Similar observations are made in the case of close-in imaging capabilities, quantified by the Inner Detectability Angle (IDA).\\ 
Constrains on DZPM design are quite relaxed at level of mask roughness (76\,nm rms in the visible spectral range). However, they can prove to be critical in the case of the mask transition width where a value better than 1\,$\mu$m is required if one wants to use our CAPM coronagraph within an exoplanet imager with a reasonable focal ratio (F less than 100) in the visible. Technological improvements of mask manufacturing are under development and some first prototypes are expected to be tested in our laboratory soon. In addition, we show that an optimization of the CAPM coronagraph with a 90\% Lyot stop is helpful for this device to be less sensitive to pupil stop misalignments.\\
Furthermore, the averaged intensity at 5\,$\lambda_0/D$ remains lower than $10^{-7}$ with our concept for a stellar angular size smaller than 0.018\,$\lambda_0/D$.\\
Central obstruction and spiders effects have not been addressed here. They will be presented in a future paper with a design study and performance analysis of the CAPM coronagraph in the framework of the E-ELT.\\
Finally, the CAPM coronagraph constitutes a very attractive concept for the planet imagers of the next decade thanks to its small inner working angle, its great ability to suppress broadband starlight and the relative simplicity of its design. The next step will now consist in manufacturing all the components of the CAPM coronagraph and performing an experimental demonstration of the concept. We are currently preparing a visible coronagraph testbed at LAM in Marseilles for this purpose. 
 

\bibliographystyle{aa}   
\bibliography{articulo03_v25}   

\end{document}